# A dust-enshrouded tidal disruption event with a resolved radio jet in a galaxy merger

S. Mattila[1,2*†], M. Pérez-Torres[3,4*†], A. Efstathiou[5], P. Mimica[6], M. Fraser[7,8], E. Kankare[9], A. Alberdi[3], M. Á. Aloy[6], T. Heikkilä[1], P.G. Jonker[10,11], P. Lundqvist[12], I. Martí-Vidal[13], W.P.S. Meikle[14], C. Romero-Cañizales[15,16], S. J. Smartt[9], S. Tsygankov[1], E. Varenius[13,17], A. Alonso-Herrero[18], M. Bondi[19], C. Fransson[12], R. Herrero-Illana[20], T. Kangas[1,21], R. Kotak[1,9], N. Ramírez-Olivencia[3], P. Väisänen[22,23], R.J. Beswick[17], D.L. Clements[14], R. Greimel[24], J. Harmanen[1], J. Kotilainen[2,1], K. Nandra[25], T. Reynolds[1], S. Ryder[26], N.A. Walton[8], K. Wiik[1], G. Östlin[12]

[1]Tuorla Observatory, Department of Physics and Astronomy, FI-20014 University of Turku, Finland.
[2]Finnish Centre for Astronomy with ESO (FINCA), FI-20014 University of Turku, Finland.
[3]Instituto de Astrofísica de Andalucía - Consejo Superior de Investigaciones Científicas (CSIC), PO Box 3004, 18008, Granada, Spain.
[4]Visiting Researcher: Departamento de Física Teórica, Facultad de Ciencias, Universidad de Zaragoza, 50019, Zaragoza, Spain.
[5]School of Sciences, European University Cyprus, Diogenes Street, Engomi, 1516 Nicosia, Cyprus.
[6]Departament d'Astronomia i Astrofisica, Universitat de València Estudi General, 46100 Burjassot, València, Spain.
[7]School of Physics, O'Brien Centre for Science North, University College Dublin, Belfield, Dublin 4, Ireland.
[8]Institute of Astronomy, University of Cambridge, Madingley Road, Cambridge, CB3 0HA, UK.
[9]Astrophysics Research Centre, School of Mathematics and Physics, Queen's University Belfast, Belfast BT7 1NN, UK.
[10]SRON, Netherlands Institute for Space Research, Sorbonnelaan 2, NL-3584 CA Utrecht, the Netherlands.
[11]Department of Astrophysics / Institute for Mathematics, Astrophysics and Particle Physics, Radboud University, P.O. Box 9010, 6500GL Nijmegen, The Netherlands.
[12]Department of Astronomy and The Oskar Klein Centre, AlbaNova University Center, Stockholm University, SE-106 91 Stockholm, Sweden.
[13]Department of Space, Earth and Environment, Chalmers University of Technology, Onsala Space Observatory, SE-439 92 Onsala, Sweden.
[14]Astrophysics Group, Blackett Laboratory, Imperial College London, Prince Consort Road, London SW7 2AZ, UK.
[15] Chinese Academy of Sciences South America Center for Astronomy, National Astronomical Observatories, Chinese Academy of Sciences, Beijing 100012, China.
[16]Núcleo de Astronomía de la Facultad de Ingeniería y Ciencias, Universidad Diego Portales,

Av. Ejército 441, 8370191 Santiago, Chile.

[17]Jodrell Bank Centre for Astrophysics, The University of Manchester, Oxford Rd, Manchester M13 9PL, UK.

[18]Centro de Astrobiología (CSIC-INTA), ESAC Campus, E-28692 Villanueva de la Cañada, Madrid, Spain.

[19]Istituto di Radioastronomia - Istituto Nazionale di Astrofisica (INAF), Bologna, via P. Gobetti 101, 40129, Bologna, Italy.

[20]European Southern Observatory, Alonso de Córdova 3107, Vitacura, Casilla 19001, Santiago de Chile, Chile.

[21]Space Telescope Science Institute, 3700 San Martin Drive, Baltimore, MD 21218, US.

[22]South African Astronomical Observatory, PO Box 9, Observatory 7935, Cape Town, South Africa.

[23]Southern African Large Telescope, PO Box 9 Observatory 7935, Cape Town, South Africa.

[24]Institute of Physics, Department for Geophysics, Astrophysics, and Meteorology, NAWI Graz, Universitätsplatz 5, 8010 Graz, Austria.

[25]Max-Planck-Institut für extraterrestrische Physik, Giessenbachstraße, 85748, Garching, Germany.

[26]Australian Astronomical Observatory, 105 Delhi Rd, North Ryde, NSW 2113, Australia.

*Correspondence to: sepmat@utu.fi, torres@iaa.es

†These authors contributed equally to the work.

**Tidal disruption events (TDEs) are transient flares produced when a star is ripped apart by the gravitational field of a supermassive black hole (SMBH). We have observed a transient source in the western nucleus of the merging galaxy pair Arp 299 that radiated >1.5×10$^{52}$ erg in the infrared and radio, but was not luminous at optical or X-ray wavelengths. We interpret this as a TDE with much of its emission re-radiated at infrared wavelengths by dust. Efficient reprocessing by dense gas and dust may explain the difference between theoretical predictions and observed luminosities of TDEs. The radio observations resolve an expanding and decelerating jet, probing the jet formation and evolution around a SMBH.**

The tidal disruption of stars by supermassive black holes (SMBH) in the nuclei of galaxies was predicted theoretically thirty years ago *(1-2)*. In a tidal disruption event (TDE), roughly half of the star's mass is ejected whereas the other half is accreted onto the SMBH, generating a bright flare that is normally detected at X-ray, ultraviolet (UV), and optical wavelengths *(3-5)*. TDEs are also expected to produce radio transients, lasting from months to years and including the formation of a relativistic jet, if a fraction of the accretion power is channelled into a relativistic outflow *(6)*. TDEs provide a means of probing central black holes in quiescent galaxies and testing scenarios of accretion onto SMBHs and jet formation *(3, 6)*.

On 2005 January 30, we discovered a bright transient in the near-infrared (IR) *(7)* coincident with the western nucleus B1 (Fig. 1) of the nearby [44.8 Mpc (*7*)] luminous infrared galaxy (LIRG) Arp 299. In galaxy mergers like Arp 299, large amounts of gas fall into the central regions, triggering a starburst. The long-term radio variability (*8*) and the IR spectral energy distribution (SED) (*9*) indicate a very high core-collapse supernova (SN) rate of ~0.3 yr$^{-1}$ within the B1 nucleus. The B1 region also harbors a Compton-thick active galactic nucleus (AGN) that has been seen directly only in hard X-rays (*10*). This is consistent with an extremely high extinction $A_V$ of ~460 magnitudes through an almost edge-on AGN torus (*11*). Galaxy mergers like Arp 299 are expected to have TDE rates several orders of magnitude higher than in field galaxies, albeit for relatively short periods of time [~$3\times10^5$ yr (*12*)].

The transient source (henceforth Arp 299-B AT1) was discovered as part of a near-IR (2.2 μm) survey for highly obscured SNe in starburst galaxies (*13*). Over the following years it became luminous at IR and radio wavelengths, but was much fainter in the optical (*7*), implying substantial extinction by interstellar dust in Arp 299. Our follow-up observations show that the nuclear outburst had a peak brightness comparable to the entire galaxy nucleus at both near-IR and radio wavelengths (Fig. 1; *8*). Based on the energetics and multi-wavelength behavior of Arp 299-B AT1 over a decade of observations (Figs. 1-3), two broad scenarios to explain its origin are plausible: (i) an event unrelated to the SMBH, such as an extremely energetic SN, or a gamma-ray burst; or (ii) accretion-induced SMBH variability, such as an AGN flare, or a TDE.

High angular resolution [100 milliarcseconds (mas)], adaptive optics assisted, near-IR imaging observations from the Gemini-North telescope (*7*) show that Arp 299-B AT1 remained stationary and coincident (within 37 mas, corresponding to ~8 pc projected distance) with the near-IR *K*-band nucleus, as seen in pre-outburst imaging from the *Hubble Space Telescope* (*HST*) (Fig. 1). Radio observations obtained with very long baseline interferometry (VLBI) constrain its position with milliarcsecond angular precision (*7*). Pre-discovery Very Long Baseline Array (VLBA) observations showed several compact sources at 2.3 GHz within the central few parsecs of the B1 nucleus, but no counterparts at higher frequencies (*14*). A new compact radio source was detected on 2005 July 17 at 8.4 GHz with the VLBA (*7*, *14*). The coincidence of the near-IR and VLBI positions, together with the appearance of the VLBI source soon after the near-IR detection and their subsequent evolution (see below; *7*), point to a common origin for both.

High angular resolution radio observations of Arp 299-B AT1 with VLBI show that the initially unresolved radio source developed a prominent extended, jet-like structure, which became evident in images taken from 2011 onwards (Fig. 1; *7*). The measured average apparent expansion speed of the forward shock of the jet is $(0.25\pm0.03)c$ between 2005 and 2015 (*7*), where *c* is the speed of light. The radio morphology, evolution and expansion velocity of Arp 299-B AT1 rule out a SN origin. Similarly, a gamma-ray burst is inconsistent with both the observed peak flux density and time to reach that peak in the radio (*15*). Therefore, the most likely explanation is that Arp 299-B AT1 is linked to an accretion event onto the SMBH. The persistent 2.3 GHz radio emission most likely corresponds to the quiescent AGN core (*7*).

The multi frequency radio light curves of Arp 299-B AT1 (Fig. 2) are well reproduced by a model (*16*) of a jet powered by accretion of part of a tidally disrupted star onto a SMBH (*7*). The jet initially moves at relativistic speeds ~0.995 *c*, but after a distance of less than ~$10^{17}$ cm (corresponding to ~760 days after the burst) it has already decelerated significantly to ~0.22 *c*, in agreement with expectations for TDE-driven jets (*6*). The apparent speed of the jet indicated by the VLBI observations, together with the non-detection of the counterjet, constrains the jet viewing angle, $\theta_{obs}$, to be within a narrow range: 25°-35° (*7*). If the jet had been launched by a pre-existing AGN, its viewing angle with respect to our line-of-sight should have been close to 90°, as the AGN torus is seen almost edge-on (*11*), and a counterjet should have also been detected (*7*). However, a radio jet associated with a TDE does not necessarily have to be perpendicular to the pre-existing AGN accretion disc (*17*). We therefore identify the observed radio jet as being launched by a TDE. No direct imaging has previously shown an expanding jet in a TDE, and its likely presence has been inferred based on unresolved radio observations only in the cases of ASASSN-14li (*18-19*), IGR J12580+0134 (*20*), and Swift J164449.3+573451 (hereafter Sw J1644+57; *21*). Our VLBI observations show a resolved, expanding radio jet in a TDE, in accordance with theoretical expectations (*6*).

The intrinsic (i.e. beaming-corrected) kinetic energy of the jet required to reproduce the radio light curves (Fig. 2) is $(1.8 \pm 0.9) \times 10^{51}$ erg (*7*), similar to the case of the relativistic TDE Sw J1644+57 (*16*). The rise of the radio emission at high frequencies in less than about 200 days and the significant delay of the lower-frequency radio emission (*7*) implies the existence of substantial external absorption, consistent with the jet being embedded in the very dense nuclear medium of the AGN, which has a constant thermal electron number density $\sim 4 \times 10^4$ cm$^{-3}$ up to a distance $6.3 \times 10^{17}$ cm from the central engine (*7*).

Observations from the ground and the *Spitzer* Space Telescope show that the IR SED of Arp 299-B AT1 and its evolution from 2005 until 2016 can be explained by a single blackbody component (Fig. 3). The blackbody radius expands from 0.04 pc to 0.13 pc between May 2005 and Jan. 2012 while its temperature cools from ~1050 K to ~750 K. The size, temperature and peak luminosity ($6 \times 10^{43}$ erg s$^{-1}$) of the IR emitting region agrees well with both theoretical predictions and observations of thermal emission from warm dust surrounding TDEs (*22-23*). Therefore, the IR SED and its evolution are consistent with absorption and re-radiation of the UV and optical light from Arp 299-B AT1 by local dust.

We modeled the IR SED of the pre- and post-outburst (734 days after the first IR detection) components of Arp 299-B1 using radiative transfer models for the emission from a starburst within the galaxy, and from a dusty torus, as predicted by the standard unified model for AGN, including also the effect of dust in the polar regions of the torus (Fig. 4, *24*). The model luminosities of the starburst and AGN dusty torus components remain constant within the uncertainties, whereas the luminosity of the polar dust component is found to increase by a factor of ~15 after the outburst, and the corresponding polar dust temperature increases from 500 to 900 K. Therefore, the observed IR SED of Arp 299-B AT1 can be most plausibly explained by re-radiation by optically thick dust clouds in the polar regions of the torus, which suffer from a relatively low foreground extinction within Arp 299-B1 (*7*).

Integrating the luminosity of Arp 299-B AT1 over the period 2005-2016 (Fig. 3) yields a total radiated energy of about $1.5 \times 10^{52}$ erg. However, a significant fraction of the total energy emitted by the transient can be expected to be scattered, absorbed, and re-radiated at substantially longer IR wavelengths by the dusty torus. We estimate that the fraction of energy that heated the polar dust was in the range 23%-78% (*7*). Thus the total radiated energy of Arp 299-B AT1 was $(1.9-6.5) \times 10^{52}$ erg, which requires a disruption of a star with a mass of about 1.9-6.5 solar masses ($M_\odot$), assuming a standard accreted fraction and radiative efficiency (*7*). Stars in this mass range can be disrupted by the $\sim 2 \times 10^7$ $M_\odot$ SMBH in Arp 299-B1 (*10, 25*). The kinetic energy of the jet is expected to be about 1% of the total rest mass energy (*6*), which agrees well with our estimated kinetic energy for the radio jet of Arp 299-B AT1 (*7*).

In addition to Arp 299-B AT1, the only other TDE candidates (although with debated nature) to have an observed radiated energy on the order of $10^{52}$ erg are ASASSN-15lh (*26-27*) and possibly transients similar to PS1-10adi (*28*). The high energy of ASASSN-15lh was originally proposed to be the result of an energetic SN (*26*), but was later explained as a tidal disruption by a high mass ($7.6 \times 10^8$ M$_\odot$), rapidly rotating black hole (*27*). In the case of PS1-10adi the large radiated energy was proposed to arise from the interaction of either an expanding TDE, or SN ejecta, with the dense medium in the nuclear environment (*28*). Arp 299-B AT1 was most plausibly the result of the disruption of a star more massive than about 2 M$_\odot$ in a very dense medium. The soft X-ray photons produced by the event were efficiently reprocessed into UV and optical photons by the dense gas, and further to IR wavelengths by dust in the nuclear environment. Efficient reprocessing of the energy might thus resolve the outstanding problem of observed luminosities of optically detected TDEs being generally lower than predicted (*29*).

The case of Arp 299-B AT1 suggests that recently formed massive stars are being accreted onto the SMBH in such environments, resulting in TDEs injecting large amounts of energy into their surroundings. However, events similar to Arp 299-B AT1 may remain hidden within dusty and dense environments, and would not be detectable by optical, UV or soft X-ray observations. The recent discovery of another TDE candidate in the nucleus of the luminous infrared galaxy IRAS F01004-2237 (*30*) yields further support for an enhanced rate of TDEs in such galaxies, which could be missed due to dust extinction. Such TDEs from relatively massive, newly formed stars might provide a large radiative feedback, especially at higher redshifts where galaxy mergers and LIRGs are more common (*31*).

**Acknowledgements**

Acknowledgements:

We thank Andrew Fabian, Talvikki Hovatta, Andrew Levan, Kari Nilsson and Claudio Ricci for useful discussions. We also thank the anonymous referees for many insightful comments which have improved the manuscript. Our findings are based mainly on observations obtained with the Spitzer Space Telescope, the European VLBI Network, the Very Long Baseline Array and Very Large Array, the Nordic Optical Telescope (NOT), and the Gemini Observatory. The Spitzer Space Telescope is operated by the Jet Propulsion Laboratory, California Institute of Technology under a contract with NASA. The European VLBI Network is a joint facility of independent European, African, Asian, and North American radio astronomy institutes. The National Radio Astronomy Observatory is a facility of the National Science Foundation operated under cooperative agreement by Associated Universities, Inc. The Nordic Optical Telescope is operated by the Nordic Optical Telescope Scientific Association at the Observatorio del Roque de los Muchachos, La Palma, Spain, of the Instituto de Astrofisica de Canarias. The Gemini Observatory is operated by the Association of Universities for Research in Astronomy, Inc., under a cooperative agreement with the NSF on behalf of the Gemini partnership: the National Science Foundation (United States), the National Research Council (Canada), CONICYT (Chile), Ministerio de Ciencia, Tecnología e Innovación Productiva (Argentina), and Ministério da Ciência, Tecnologia e Inovação (Brazil). Figure 1 image credit: NASA, ESA, the Hubble Heritage Team (STScI/AURA)-ESA/Hubble Collaboration and A. Evans (University of Virginia, Charlottesville/NRAO/Stony Brook University).

Funding:
SM acknowledges financial support from the Academy of Finland (project: 8120503). The research leading to these results has received funding from the European Commission Seventh Framework Programme (FP/2007-2013) under grant agreement numbers 227290, 283393 (RadioNet3) and 60725 (HELP). AA, MPT, NRO and RHI acknowledge support from the Spanish MINECO through grants AYA2012-38491-C02-02 and AYA2015-63939-C2-1-P. PGJ acknowledges support from European Research Council Consolidator Grant 647208. C.R.-C. acknowledges support by the Ministry of Economy, Development, and Tourism's Millennium Science Initiative through grant IC120009, awarded to The Millennium Institute of Astrophysics, MAS, Chile and from CONICYT through FONDECYT grant 3150238 and China-CONICYT fund CAS160313. PM and MAA acknowledge the support from the ERC research grant CAMAP-250276, and the partial support from the Spanish MINECO grant AYA2015-66889-C2-1P and the local Valencia government grant PROMETEO-II-2014-069. MF acknowledges support from a Science Foundation Ireland - Royal Society University Research Fellowship. DC acknowledges the support from the grants ST/G001901/1, ST/J001368/1, ST/K001051/1, ST/N000838/1. PV acknowledges support from the National Research Foundation of South Africa. JH acknowledges financial support from the Finnish Cultural Foundation and the Vilho, Yrjö and Kalle Väisälä Foundation.

Author contributions:
SM and MPT co-led the writing of the manuscript, the data analysis and physical interpretation. AE modeled the IR SED and contributed to the physical interpretation and text. PM and MAA modeled the radio light curves and contributed to the physical interpretation and text. MF analyzed the HST data and contributed to the physical interpretation and text. EK contributed to the observations and analysis of the near-IR data, the physical interpretation and text. AA, CRC, and IMV contributed to the analysis and interpretation of the radio data, and text. EV, MB, RHI, NRO, RB and KW contributed to the analysis and interpretation of the radio data. TH and ST analyzed the X-ray data. PJ and SS contributed to the physical interpretation and text. PL and CF contributed to the physical interpretation. AAH, WM, RK and PV contributed to the analysis and physical interpretation of the infrared data. JH, TK and TR contributed to the observations and analysis of the near-IR data. DC, JK, KN, RG, SR, NW and GÖ contributed data. All co-authors contributed with comments to the text.

Competing interests:
We declare that none of the authors have any competing interest.

Data and materials availability:
The raw observations used in this publication are available from the Spitzer Heritage Archive at http://sha.ipac.caltech.edu/applications/Spitzer/SHA/ (Proposal IDs: 32, 108, 60142, 80105, 90031, 90157, 10086, 11076), from the EVN data archive at http://archive.jive.nl/scripts/listarch.php (proposal IDs EP063, EP068, EP075, EP087, GP053), the NRAO data archive at https://archive.nrao.edu/archive/advquery.jsp, from the NRAO data archive at https://archive.nrao.edu/archive/advquery.jsp (proposal IDs: BPU027, BP202, AC0749), the NOT data archive at http://www.not.iac.es/archive/, the Gemini Observatory Archive at https://archive.gemini.edu/searchform (programs: GN-2008B-Q-32, GN-2009A-Q-12, GN-2009B-Q-23, GN-2010A-Q-40, GN-2011A-Q-48 and GN-2011B-Q-73), the Hubble Legacy Archive at https://hla.stsci.edu/hlaview.html, the Chandra Data Archive at http://cxc.harvard.edu/cda/ (OBSIDs 1641, 6227, 15077 and 15619), the XMM-Newton Science Archive at http://nxsa.esac.esa.int/nxsa-web/ (ObsId 0679381101), the Isaac Newton Group Archive at http://casu.ast.cam.ac.uk/casuadc/ingarch/query, the United Kingdom Infrared Telescope Archive at http://casu.ast.cam.ac.uk/casuclient/ukirt_arch/, The radiative transfer models used in this paper are part of the CYGNUS project with the model grids available at http://ahpc.euc.ac.cy/index.php/resources/cygnus. The results of the hydrodynamic and radiative simulations used for modeling the radio light curves are available at https://www.uv.es/mimica/doc, as are the data and Python code used to produce Fig. 2A and Fig. S6. The Python code used for determining the allowed values for the viewing angle of the radio jet and producing Fig. S7 is available at https://github.com/mapereztorres/rad-trans-theta. Full details of all data and software used in this paper are given in the supplementary material.

**Supplementary Materials**
www.sciencemag.org
Materials and Methods
Figs. S1 to S7
Tables S1 to S8
References (33-99)

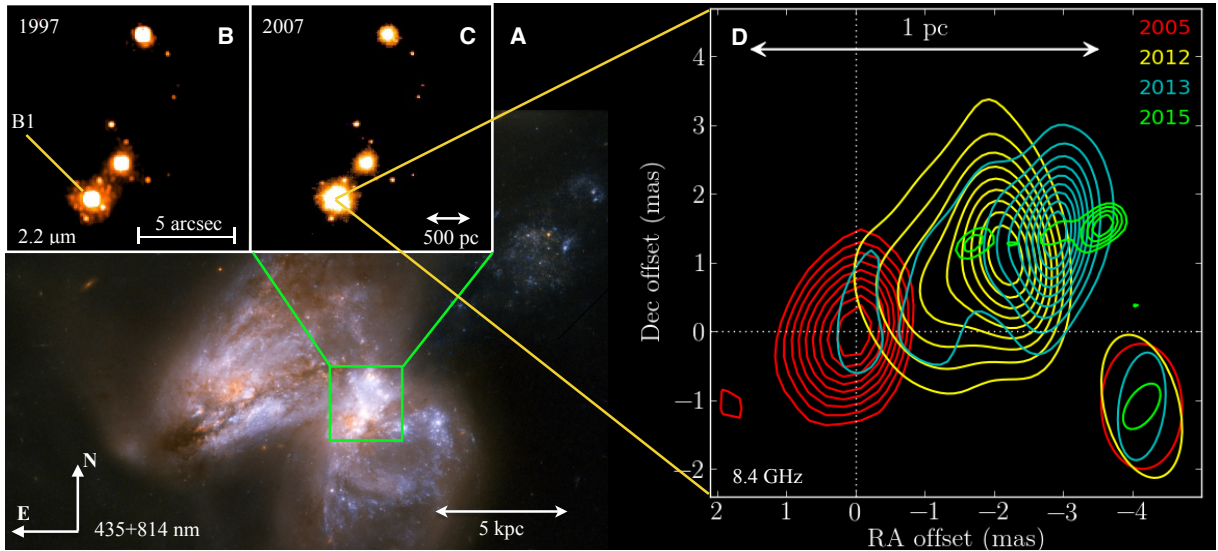

**Figure 1. The transient Arp 299-B AT1 and its host galaxy Arp 299.** (**A**) A color-composite optical image from the Hubble Space Telescope (*HST*), with high resolution, 12.5×13 arcsecond size near-infrared 2.2 μm images (insets **B** and **C**) showing the brightening of the B1 nucleus (*7*). (**D**) The evolution of the radio morphology as imaged with Very Long Baseline Interferometry (VLBI) at 8.4 GHz (7×7 milliarcsecond region centered at the 8.4 GHz peak position in 2005, RA = 11h28m30.9875529s, Dec = 58°33′40″.783601 (J2000.0), indicated by the dotted lines). The VLBI images are aligned with an astrometric precision better than 50 μas. The initially unresolved radio source develops into a resolved jet structure a few years after the explosion, with the center of the radio emission moving westward with time (*7*). The radio beam size for each epoch is indicated in the lower-right corner.

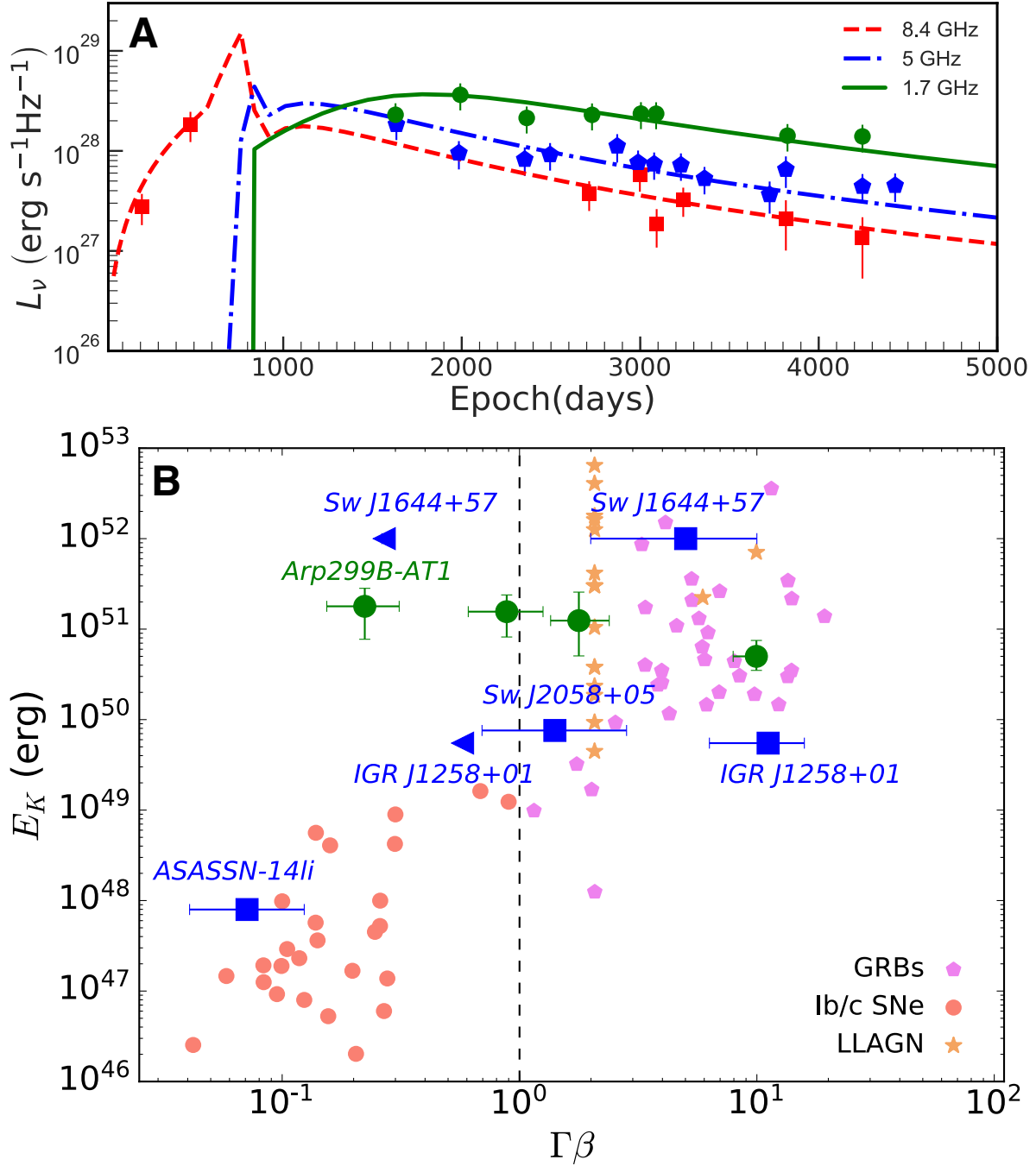

**Figure 2. Radio properties of Arp 299-B AT1.** (A) Radio luminosity evolution of Arp 299-B AT1 at 1.7 (circles), 5.0 (pentagons) and 8.4 GHz (squares) spanning more than 12.1 years of observations, along with modeled radio light curves, using hydrodynamic and radiative simulations for a tidal disruption event (TDE) launched jet *(16)*. The day zero corresponds to 2004 Dec. 21.6. (B) Intrinsic (beaming-corrected) jet kinetic energy, $E_K$, versus outflow speed ($\Gamma\beta$, where $\Gamma = (1-\beta^2)^{-1/2}$ is the bulk Lorentz factor of the outflow and $\beta = v/c$), from radio observations of gamma-ray bursts (GRB), supernovae (SNe), low-luminosity active galactic nuclei (LLAGN), and TDEs *(4, 16, 19-21, 32)*. The large circles show, from right to left, the inferred loci for Arp 299-B AT1 at four different epochs in the observer's frame: just after the jet is launched by the TDE, and ~1, ~12, and ~760 days thereafter. For the LLAGN sample, we have assumed a constant jet kinetic power over 10 yr. The triangles indicate upper limits for the expansion speed of IGR J1258+01 *(20)* and Sw J1644+57 *(21)*.

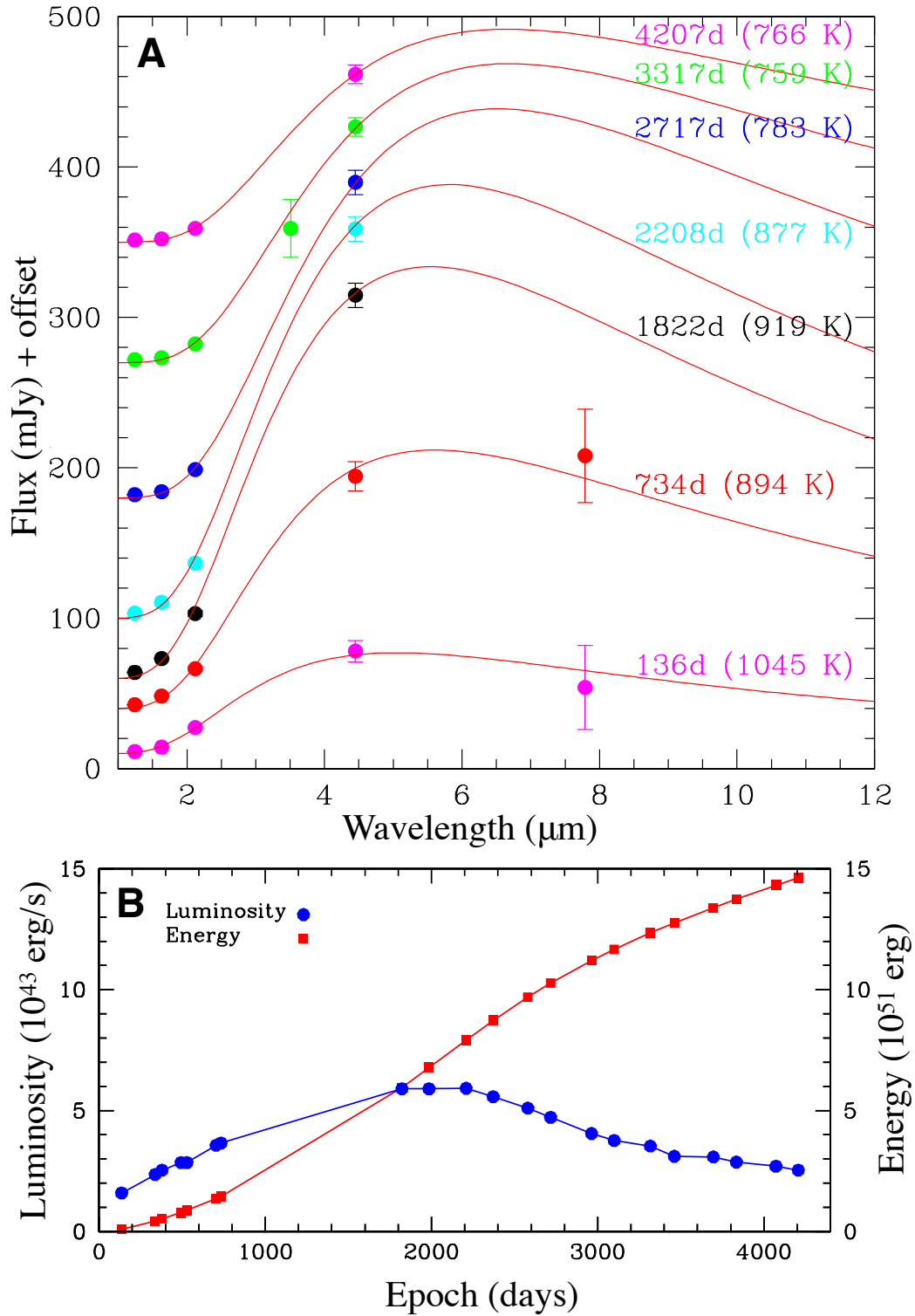

Figure 3. Infrared properties of Arp 299-B AT1. (A) Evolution of the observed infrared spectral energy distribution (points) shown together with blackbody fits between 136 and 4207 days after the first infrared detection on 2004 Dec. 21.6 (*7*). Over this period the blackbody temperature decreased from about 1050 to 750 K while the blackbody radius increased from 0.04 to 0.13 pc. (B) The evolution of the integrated blackbody luminosity (blue circles) and cumulative radiated energy (red squares). The observed radiated energy by day 4207 was about $1.5 \times 10^{52}$ erg.

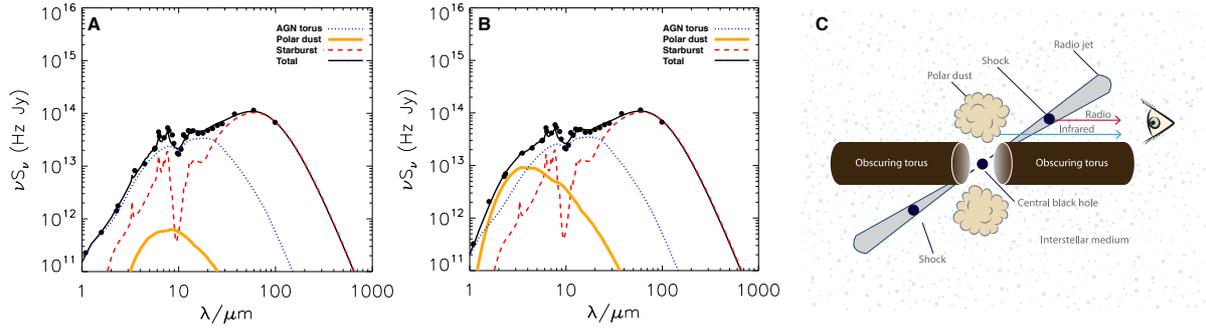

**Figure 4. Model for the observed properties of Arp 299-B AT1.** Best-fit models for the spectral energy distribution of the Arp 299 nucleus B at **(A)** pre-outburst **(A)** and **(B)** post-outburst (734 days after the first mid-infrared detection). The models include a starburst component (dashed line), an active galactic nucleus (AGN) dusty torus (dotted line), and a polar dust component (thick solid line) (*7*). The sum of these components is shown as a thin solid line. In **(B)** most of the model parameters were fixed, whilst the temperature of the polar dust varied from 500 K in the pre-outburst case to 900 K in the post-outburst case. This yields a covering factor of the polar dust of 23%-78%, implying that the total radiated energy is ~(1.9-6.5) ×$10^{52}$ erg. **(C)** Schematic diagram (not to scale) showing the geometry of the emitting and absorbing regions (*7*). The tidal disruption event generates prominent X-ray, ultraviolet and optical emission. However, the direct line-of-sight to the central black hole is obscured by the dusty torus, which is opaque from soft X-rays to infrared wavelengths. The polar dust re-radiates in the infrared a fraction of the total energy emitted by the event. The transient radio emission originates from a relativistic jet launched by the tidal disruption of a star.

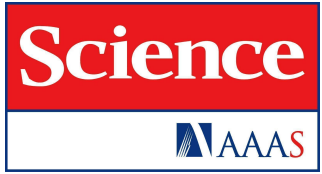

Supplementary Materials for

A dust-enshrouded tidal disruption event with a resolved radio jet in a galaxy merger

Seppo Mattila, Miguel Pérez-Torres et al.

correspondence to: sepmat@utu.fi, torres@iaa.es

**This PDF file includes:**

Materials and Methods
Figs. S1 to S7
Tables S1 to S8

# Materials and Methods

## The host galaxy of the transient Arp 299-B AT1

The merging pair of galaxies Arp 299 (Fig. 1) at a 44.8 Mpc distance (*33*; corrected to the cosmic microwave background reference frame; $H_0 = 73$ km s$^{-1}$ Mpc$^{-1}$) is one of the brightest nearby LIRGs with an IR luminosity $L_{IR} \sim 7 \times 10^{11}$ $L_\odot$ (*34*; $L_\odot = 3.833 \times 10^{26}$ W). The system consists of two galaxies, Arp 299-A and Arp 299-B, and the merger is thought to have begun around 750 Myr ago (*35*). Arp 299 has an estimated total star formation rate of ~150-200 $M_\odot$ yr$^{-1}$ and a total core-collapse SN rate of ~1.5-2 SNe yr$^{-1}$ (*9*). The system has continued to experience star formation triggered over the last ~50 Myr (*11*). Arp 299-B hosts two prominent nuclei, B1 and B2. A strongly obscured AGN was discovered in the B1 nucleus using data in the hard X-ray band collected by the BeppoSAX observatory in 2001 (*36*).

## Optical and infrared observations and data analysis

We monitored Arp 299 in the near-IR *Ks*-band from 2002 Jan. to 2005 Jan. as part of a program to search for obscured SNe in the nuclear regions of nearby starburst galaxies (*37-38*), using the Isaac Newton Group Red Imaging Device (INGRID; *39*) and Long-slit Intermediate Resolution Infrared Spectrograph (LIRIS; *40*) instruments on the 4.2-m William Herschel Telescope (WHT). We reduced the images and compared them to a reference image using standard Image Reduction and Analysis Facility (IRAF; *41*) routines and the ISIS2.2 image subtraction package (*42-43*), to produce difference images. This procedure revealed a strong outburst in the B1 nucleus of Arp 299 in the *Ks*-band images obtained on 2005 Jan. 30 (*13*). The type II SN2005U in Arp 299 was also discovered in the same images (*44*).

To test whether the transient Arp 299-B AT1 was detectable in the optical, we obtained *r* and *i* band imaging with the Wide Field Camera (WFC) on the 2.5-m Isaac Newton Telescope (INT) in 2005 and 2006. For the photometric calibration of the images we used the *r* and *i* band magnitudes (in the AB system; *45*) of six field stars available from the Sloan Digital Sky Survey (SDSS; *46*). We compared these images with pre-outburst images obtained with the INT in 2002 using the same filters and instrument setup. The post-outburst images were aligned with the pre-outburst images and image subtraction was carried out using ISIS2.2. However, our *r* and *i*-band observations did not yield an optical detection of Arp 299-B AT1, and we estimate an *i*-band limiting magnitude of ~17.5 for a point source at the position of the *K*-band nucleus B1 in the 2005 Mar. 24 epoch.

Arp 299 has been observed on multiple occasions by the *HST*. We used pipeline-reduced and drizzled images from the Hubble Legacy Archive obtained using the *F814W* filter (close to *I*-band) and several different instruments: in 1998 Oct. 25.1 Wide-Field Planetary Camera 2 (WFPC2; *47*) with Planetary Camera (PC), in 2000.0 July 11.2 WFPC2, in 2001 Mar. 16.4 WFPC2 with PC, in 2001 May 31.1 WFPC2 with PC, in 2006 Mar. 19.3 Advanced Camera for Surveys (ACS; *48*) with Wide-Field Channel (WFC) and 2010 Jun. 24.9 Wide-Field Camera 3 (WFC3; *49*) with Ultraviolet-Visible channel (UVIS). In cases where multiple consecutive exposures were taken using the same filter, we combined them into a single deep image. We derived a geometric transformation between each of the earlier images and the 2010 image, using 17 point sources and allowing for rotation, an independent scaling in x and y, and a polynomial term. Once all the *F814W* frames were in the same pixel coordinates, we performed difference imaging between the 2010 frame and all other images. As the images were taken with a range of instruments, with different pixel scales and point spread functions (PSFs), we used the HOTPANTS package (*50*) to convolve the images and match their PSFs before scaling them to a common flux level and subtracting them.

The best subtraction, as judged by the absence of residuals and a clean uniform background across the difference image, was obtained when subtracting the 2006 ACS image from the 2010 WFC3 image, as shown in Fig. S1. In this subtraction, we see two sources close to Arp 299-B1. The first source corresponds to the nucleus, and is coincident (within 7 mas) with the source seen in *K*-band difference imaging. This source is brighter in 2010 than in 2006. The second source is offset by 0.2″ to the East of the nucleus, and is brighter in 2006 compared to 2010. We performed aperture photometry on both sources in the difference image using a 0.1″ aperture (Fig. S1), and set the background flux to zero. We corrected our measured fluxes to an infinite aperture but did not correct for charge transfer efficiency losses, as these should be small (~0.01 mag) given the date of the observations and the high background levels (*51-52*). We obtain a magnitude of F814W = 19.76±0.01 for the source coincident with the nucleus, and F814W = 22.07±0.01 for the offset source in 2006. The photometric uncertainties correspond solely to the Charge-Coupled Device (CCD) noise although there is likely a larger systematic error due to the difference imaging.

We obtained ground-based near-IR (*J, H, Ks*-band) imaging follow-up of Arp 299-B AT1 with LIRIS on the WHT, the 3.8-m United Kingdom Infrared Telescope (UKIRT) and UKIRT Fast-Track Imager (UFTI; *53*), NOTCam (*54*) on the 2.6-m Nordic Optical Telescope (NOT), and The Instrument Formerly Known As Mosaic (TIFKAM) on the 2.4-m Hiltner telescope. Our near-IR imaging follow-up thus covers epochs between 2005 Jan. and 2016 Dec. The near-IR images were reduced and the quiescent host contribution subtracted following the procedures detailed above and in (*55*). A WHT/LIRIS image obtained on 2004 Jun. 5 was used as the reference image in *K*-band and images from National Science Foundation camera (NSFCAM; *56*) on the 3.0-m NASA Infrared Telescope Facility (IRTF) obtained on 1998 Mar. 2 were used as the reference images in *J*- and *H*-bands (*57*).

For the astrometric calibration the LIRIS *Ks*-band image was registered with a Two Micron All-Sky Survey (2MASS; *58*) *Ks*-band image with absolute astrometric information available. The centroid coordinates for Arp 299-B AT1 were then measured in the subtracted image at the epoch of discovery, yielding a right ascension (RA) of 11h28m30.97s and declination (Dec) of +58°33′40.9″ (J2000.0 equinox) with an estimated uncertainty of 0.5″ in each coordinate. For the photometric calibration of the reference images we used the *J, H,* and *K* band magnitudes (on the Vega system) of five field stars surrounding Arp 299 available from 2MASS (*55*) and transformed the calibration to the IRTF images using NOTCam images with a larger field of view. We carried our aperture photometry of Arp 299-B AT1 in the subtracted images using the Starlink package GAIA (*59*). An aperture radius of 3″ was used. The photometric errors are dominated by the systematic uncertainties in the calibration and uncertainties arising from the image subtraction method. The former was estimated as the standard error of the mean of the zero point magnitudes obtained with the different field stars and the latter from the scatter between the consecutive light curve points during a phase of no apparent light curve evolution. We estimate a photometric uncertainty of 4%, including both of these effects. See Table S1 for the log of near-IR observations and the photometric results.

We observed Arp 299 using the Gemini-North Telescope (Gemini-N) for a total of ten epochs (between 2007 Mar. 6 and 2012 Jan. 29; PIs: S. Smartt and S. Ryder) with the Near-

Infrared Imager (NIRI) and the ALTAIR laser guide star adaptive optics (AO) system (0.022″/pix, FWHM ~ 0.1″). The individual NIRI images were reduced with IRAF using the NIRI package in IRAF and registered and co-added using a bright star as a reference (for details see *55*). *HST* Near Infrared Camera and Multi-Object Spectrometer (NICMOS, *60*) observations (*61*) with the NIC2 camera (0.076″/pix, FWHM ~0.1″) and *F222M* filter on 1997 Nov. 5 provided a reference image with the nucleus B1 at the quiescent level.

We derived a geometric transformation between each of the Gemini-N images and the *HST* image using 9 point-like sources common between the images and a transformation including x and y shifts, a scaling factor and a rotation angle. We made use of this transformation to align the *HST* reference image to each Gemini-N image prior to subtracting between these two using the ISIS2.2 package. The centroid coordinates of the outburst were then measured in the subtracted images. The r.m.s. uncertainties of the geometric transformation and the calculated offsets between the coordinates of the quiescent nucleus and the outburst are listed in Table S2. We found the RA of the outburst in the nucleus B1 to be coincident (within the 1σ uncertainty) with the quiescent nucleus in all the epochs (also see Fig. 1). In Dec, there are offsets of up to 37 mas for the individual epochs between the outburst and the quiescent nucleus, although the offsets are not statistically significant (about 2σ). It is also likely that our method underestimates the astrometric uncertainties given the different instruments used for the pre- and post-outburst observations. Therefore, we conclude that Arp 299-B AT1 is coincident with the quiescent nucleus B1 and we see no evidence for any systematic shift in the nuclear position between the epochs.

Mid-IR follow-up observations of the outburst were serendipitously obtained from the *Spitzer* Space Telescope. Arp 299 was monitored at 4.5 μm and 8.0 μm with the Infrared Array Camera (IRAC, *62*) instrument from 2004 May every ~6 months over 10 epochs as a part of a search for obscured SNe (PI: C. Lawrence). We continued the mid-IR monitoring of the outburst as a part of the Spitzer warm mission at 4.5 μm (PI: S. Mattila) for 14 epochs in 2009-2016. We also included archival 3.6 μm IRAC observations obtained in 2013-2015 (PI: O. Fox). IRAC observations at 3.6, 4.5, and 8.0 μm obtained in 2003 (PI: G. Fazio) provided reference images of Arp 299-B1 at the quiescent level. For the log of mid-IR observations see Table S3.

The 4.5 and 8.0 μm IRAC observations comprised four exposures per dithering position, the first with an exposure time of 0.6 sec, and the other three each with an exposure time of 12 sec. Although the nucleus B1 was badly saturated in all 12 sec frames, the 0.6 sec frames were useful for accurate flux measurements. We used the post-basic calibrated data (PBCD) product short exposure mosaics throughout (pipeline versions S18.25, 19.1 and 19.2). In the case of the warm mission 3.6 μm observations only one short exposure had been obtained per epoch and the basic calibrated data (BCD) products were used. We carried out aperture photometry of Arp 299-B AT1 in the subtracted images using GAIA. A circular aperture of radius 3.66″ was used and the photometry was corrected to an infinite aperture (*63*). The sky was measured using a clipped mean sky estimator and concentric annuli outside the galaxy. For nucleus B1, which is a very bright source, we assume that the photometric uncertainty is dominated by the Spitzer absolute flux calibration. We adopt 3% errors for all the cryogenic observations and post-cryogenic PBCD mosaics processed with a pipeline version earlier than S19.2, 2% errors for the post-cryogenic mosaics processed with the S19.2 pipeline and 10% errors for the 3.6 μm measurements made on the individual BCD frames (*63*). The photometric results are presented in Table S3 (before subtracting the quiescent nuclear flux) and Fig. S2 (after subtracting the quiescent flux based on the first epoch of 3.6 μm observation and the first three epochs of 4.5 μm and 8.0 μm observations). The epoch when the mid-IR light curves were observed to start rising, 2004 Dec. 21.6 corresponding to the Julian Date (JD) 2453361.1, was adopted as day 0 (Julian Date is used by astronomers and corresponds to the number of days since year 4713 BC plus the fraction of a day since the preceding noon in Universal Time).

**Radio observations and data analysis**

We imaged the central few parsecs of Arp 299-B1 using radio interferometry data taken with the European VLBI Network (EVN) (PI: M. Pérez-Torres), the VLBA (PI: M. Pérez-Torres) and global VLBI, which included both the EVN and the VLBA (PI: M. Pérez-Torres) at 1.5, 1.7, 2.3, 5.0 and 8.4 GHz. These observations were part of a VLBI monitoring program aimed at characterizing the population of compact radio objects in the central starburst regions and estimating the (radio) SN rates therein *(64, 65)*. We also used publicly available archival data taken with the Very Large Array (VLA) at multiple frequencies, as well as VLBA data at 2.3 and 8.4 GHz. (See Table S4 for a log of the observations.)

The EVN and global VLBI data were correlated at the EVN MkIV Data Processor at the Joint Institute for VLBI in Europe (JIVE) in The Netherlands, while the VLBA data were correlated at the Array Operations Center in Socorro (USA). We chose time and frequency intervals small enough to guarantee negligible time and bandwidth-smearing effects in the data. We phase-referenced our observations to the VLBI calibrator NVSS J112813+592515. We used the strong source 4C 39.25 as fringe finder and bandpass calibrator and, when this was not possible, we used J1128+5925 for those purposes. The data were later correlated at the different nuclei of Arp 299, which included our target source, the nucleus Arp 299-B1. We performed standard a-priori gain calibration using the measured gains and system temperatures of each antenna. This calibration, as well as the data inspection and flagging, were done within the National Radio Astronomy Observatory (NRAO) Astronomical Image Processing System (*AIPS*; *66*) assisted with ParselTongue *(67)*, which we used as an interface to *AIPS*. We also corrected for ionosphere effects (of particular relevance at 1.7 and 2.3 GHz) and source-structure effects of the phase-reference source at all three frequencies, following the same procedures described in *(64-65)*.

We complemented our datasets with archival VLA and VLBA + Green Bank Telescope (GBT) data taken at 2.3 and 8.4 GHz, as well as with VLA data at 8.4 and 22.5 GHz. The angular resolution of the VLA is much poorer than that provided by VLBI arrays, so the VLA cannot resolve the compact radio emission. Fortunately the nuclear region of Arp 299-B1 is faint enough at high frequencies that the strong burst in 2005 could be detected with the VLA [*(8)* and Table S4]. Also, the VLA flux density measurements could contain contributions from surrounding sources, so we have assumed a 30% calibration error for those measurements. The archival NRAO VLBA+GBT observations at 2.3 and 8.4 GHz included a pre-discovery (Jan. 2005) and a post-discovery epoch (Jul. 2005). These data were published in *(14)*. We independently re-calibrated and re-analyzed these data, following the same data reduction procedures described above. We also analyzed archival VLBA+GBT data prior to 2005. However, these data had severe calibration problems, which prevented us from detecting any source in the Arp 299 system, and therefore were discarded.

We carried out all imaging and deconvolution procedures within *AIPS*, using task IMAGR with Briggs weighting *(68)* and a ROBUST parameter of 0.5 (Figs. 1 and S4). This scheme was a good compromise to combine a very high angular resolution and a very low off-source r.m.s. value in our images. For the archival VLBA+GBT data, we used instead a ROBUST parameter of 0 for comparison purposes with the imaging in *(14)* (see also Fig. S3). The off-source r.m.s. value varied significantly from epoch to epoch, but was significantly lower than the calibration uncertainty, which was typically 10%, depending on the array and observing frequency. We list in Table S4 the multi-frequency radio interferometric observations of Arp 299-B AT1, spanning more than 12 years of observations for a total of more than 30 epochs of milliarcsecond angular resolution radio observations, plus lower angular resolution observations taken with the VLA at 8.4 and 22.5 GHz.

## X-ray observations and data analysis

In the energy range below ~10 keV, Arp 299 has been observed with the *Chandra* and *XMM-Newton* observatories several times during 2002-17 resulting in 5 publicly available datasets including our own observations obtained through DDT requests (PI: P. Nandra and PI: M. Pérez-Torres, respectively; see Table S5). Despite the strong absorption and existence of additional sources of emission in this energy band we can use these observations to constrain the AGN variability on ~year time scales. We reanalyzed all *Chandra* and *XMM-Newton* data in a consistent way.

We extracted the X-ray spectrum of Arp 299-B1 from the *Chandra* data for three epochs: 2001, 2005 and 2013 (where the last epoch consists of observation done on consecutive days in 2013). First, all the data were aligned, and all standard corrections were applied using CIAO 4.8 *(69)* and CALDB 4.7.0 *(70)*. The spectra were extracted using CIAO's specextract tool from a circular source aperture at the Arp 299-B1 position with a radius of 1.5″. The background was measured from an empty region outside the galaxy. The spectra from observations in 2013 were combined. All the extracted spectra were grouped so that each spectral bin had a minimum of one count. The spectra were analyzed using XSPEC 12.9.0i *(71)*, using C-statistics.

Only channels in the energy range 0.3-8.0 keV were included for the purposes of fitting. LIRGs such as Arp 299 are well known X-ray emitters from several different sources (thermal emission of hot gas, X-ray binaries associated with the starburst, AGN). Furthermore, due to the very strong absorption inferred from the spectral analysis (corresponding to a hydrogen column density of $N_H = 3 \times 10^{24}$ cm$^{-2}$) direct emission from the AGN is detectable only at energies above 10 keV. We adopted the best fitting model from *(72)* as initial values, and fitted this model for the combined spectrum from 2013 allowing all parameters to vary freely. All the emission components were absorbed via photoelectric absorption (we used the XSPEC model phabs(apec+pcfabs(po+gau))). For the AGN, we adopted the spectral model *(72)* consisting of a power-law with additional Gaussian line components to represent the ~6.4 keV iron emission line seen through a partial covering absorber plus a thermal plasma component. The results of this fit were consistent within the errors with the values obtained in *(72)*. We then took our best-fit values for this observation as starting values for the spectra extracted from the 2001 and 2005 observations. We fitted the same model for both of these spectra separately, but only allowing the power-law photon index and the normalization parameters of all components (thermal plasma, power-law and Gaussian) to vary freely. We also measured, for all three spectra, the flux of the fully absorbed model as well as the unabsorbed flux of the power-law component (both in the 0.3-8.0 keV energy range) and the equivalent width of the iron line (see Table S5). We find no evidence for statistically significant variability in the full model flux, power-law component flux, photon index or equivalent width of the iron line.

*XMM-Newton* observed Arp 299 twice, in May 2001 and Dec. 2011. However, the poorer angular resolution than *Chandra* did not permit us to isolate photons from the Arp 299-B1 nucleus alone. For these data the source aperture had a radius of 15″ and therefore collected almost all photons from the western component of Arp 299. The background region was located outside the galaxy and contained neither point sources nor a diffuse thermal component. The fitting model was taken to be exactly the same as for the *Chandra* data. However, taking into account the very different source apertures, the absolute values of the resulting fluxes obtained with *XMM-Newton* and *Chandra* observatories differ by a factor of a few owing to a contribution from the unresolved high-mass X-ray binaries residing in the central parts of the western component of Arp 299. Nevertheless, we can compare the fluxes derived from the two *XMM-Newton* observations well before and after the discovery of Arp 299-B AT1. We conclude that the soft X-ray flux from the B1 nucleus has remained constant within the uncertainties.

## Proper motion of Arp 299-B AT1

Our VLBI observations allowed us to accurately determine the average proper motion of Arp 299-B AT1. We show in Fig. S5 model fits to the proper motion in both the RA and Dec directions, over a time span of more than 12 yr. We used only the highest angular resolution VLBI data (at 5.0 and 8.4 GHz) up to 4000 days for fitting the proper motion of the jet, as the lower frequency data have a much lower angular resolution, especially in the North-South direction. The fitted angular velocities are $\mu_{RA} \sim -(0.322\pm0.032)$ mas yr$^{-1}$ and $\mu_{Dec} \sim (0.151\pm0.015)$ mas yr$^{-1}$, which correspond to an average apparent transverse linear velocity of $\beta_{app} = (0.25\pm0.03)$.

The apparent change in the proper motion in the Dec coordinate after 4000 days occurs when observations switched from the EVN/Global VLBI observations to VLBA-only observations, so the change in proper motion beyond ~4000 days is likely an artifact caused by the poorer angular resolution of the VLBA synthesized beam. We emphasize that the proper motion estimates found in the paper are not affected by this potential change at late times.

## The nature of Arp 299-B AT1

Pre-discovery VLBA+GBT observations of Arp 299-B AT1 showed four compact sources within the B1 nucleus at the frequency of 2.3 GHz, but no counterparts at 8.4 GHz [see *(14)*, their figure 4]. We show in Fig. S3-4 the resulting images of our reanalysis of the 8.4 GHz VLBA+GBT observations in Jan. 2005 and Jul. 2005. The 8.4 GHz observations in Jan. 2005 do not show evidence for radio emission above the noise, while the observations in Jul. 2005 show clear evidence for a new, bright radio source (at RA = 11h28m30.9875529s, Dec = 58°33′40.783601″; J2000.0.0) with a peak luminosity at 8.4 GHz of $\nu L_\nu \sim 1.6 \times 10^{38}$ erg s$^{-1}$.

We monitored the evolution of the flux density and radio morphology of Arp 299-B AT1 using VLBI observations. The high flux density and compactness of the radio emission from Arp 299-B AT1 implies a non-thermal origin. Initially, the source appeared unresolved (Fig. S3, right), and therefore its compactness and high flux density could have been explained by a number of scenarios, either extrinsic to the SMBH (e.g., an extremely energetic SN, or a GRB), or intrinsic (e.g., accretion-induced SMBH variability, such as an AGN flare, or a TDE). Our VLBI monitoring shows that Arp 299-B AT1 developed a jet-like morphology, indicating an apparent expansion speed $\beta_{app}$ = 0.25±0.03 (Figs. 1 and S4). Long GRBs occur preferentially in low-metallicity environments (e.g. *73*) whereas Arp 299 has a roughly solar metallicity (*74*). While GRBs may exhibit jet-like features, the time-to-peak (a few days to a few tens of days), 8.4 GHz peak radio luminosity and flux density evolution of both short- and long-GRBs (*15*) are at odds with the observed radio behavior of Arp 299-B AT1. A TDE from an intermediate mass black hole (IMBH) is an unlikely scenario, given the very small number of known IMBH candidates and the fact that the location of the transient coincides with the dynamical mass center of Arp 299-B1 (as traced by the 2.2 μm light). In addition, the IMBH scenario would also require the disruption of a subsolar mass white dwarf (*25*) which is not consistent with our observations (see below). We therefore conclude that Arp 299-B AT1 is an event related to the SMBH.

The compact 2.3 GHz emission detected with VLBI in January 2005 very close to the VLBI position of Arp 299-B AT1 in July 2005 (Fig. S3), together with the long-term evolving radio morphology, suggests that this 2.3 GHz emission pinpoints the location of the AGN in Arp 299-B1. This persistent 2.3 GHz emission could in principle be explained by two alternative scenarios: (i) the AGN core itself; or (ii) a stationary component of the jet structure very close to the base of the AGN. In the AGN-core scenario, the initial 8.4 GHz compact emission observed in Jul. 2005 would correspond to the emission of a single plasmoid from the AGN core, which is shifted at 8.4 GHz with respect to the 2.3 GHz peak (Fig. S3) because of the well-known core-shift (*75*), whereby the peaks of emission at different frequencies are shifted due to opacity effects. The observed shift would correspond to a linear offset of (0.35±0.11) pc, which is not unexpected for AGN (*76*). An alternative scenario would be that of the 2.3 GHz persistent radio emission being due to a stationary component of the pre-existing AGN jet in Arp 299-B1. However, since the AGN torus is seen almost edge-on (*11*) and the activity of a normal AGN jet in Arp 299-B1 would be along the polar direction, essentially no beamed emission from such a pre-existing jet would be detected. Therefore, the observed behavior of the compact VLBI radio emission appears to be at odds with the 2.3 GHz radio emission being from a stationary component, and we assume hereafter that the 2.3 GHz VLBI peak observed in 2005 corresponds to the core of the AGN in Arp 299-B1. We emphasize that in either case the 2.3 GHz peak would correspond to pre-existing emission of the AGN.

We conclude that Arp 299-B AT1 is the result of a sudden change in the accretion rate onto the SMBH induced by the tidal disruption of a star. Our high-angular resolution VLBI imaging at 8.4 GHz shows (Figs. 1 and S4) that the radio emission at the initial location has faded completely a few years later, while the radio jet has been evolving and expanding over about 10 years. Therefore, the pre-existing low-luminosity AGN, probably being fed at a very low accretion rate, returned to its steady state shortly after the TDE flare while the TDE-launched jet continued expanding into a dense surrounding medium, and therefore decelerated. The average expansion speed of Arp 299-B AT1 requires a substantial deceleration of the radio jet over a 12 year period, which is at odds with a jet launched by a low-luminosity AGN (*32*; see also Fig. 2B).

The 8.4 GHz VLBI burst in Arp 299-B AT1 and its observed radio jet morphology and evolution are most naturally explained by a TDE. We therefore modeled the data in Table S4 using the same techniques as outlined in (*16, 77-78*), which we briefly outline here. We first simulated the hydrodynamical evolution using the MRGENESIS code (*79-80*). The radio emission was then modeled using the SPEV code (77-78). The (radio) synchrotron emission arises from relativistic electrons that follow a power-law in energy, $N(E) \sim E^{-p}$, where $p$ is the index of the injected power-law of relativistic electrons. The energy feeding the relativistic electrons and the magnetic field comes from the post-shock thermal energy: a fraction, $\varepsilon_e$, of the post-shock thermal energy goes to the relativistic electrons, and a fraction $\varepsilon_B$, to the magnetic field. The values of $\varepsilon_e$ and $\varepsilon_B$ are uncertain, but are unlikely to be much larger than ~0.1 (e.g. *16*). Additionally, the code in (*16, 77-78*) allows the specification of a fraction $\zeta_e$ of the available electrons to be accelerated at the shock. Here, we adopted $\zeta_e = 1$. Fig. 2A shows the best-fit modeled radio light curves, which are obtained for $p = 3.0$, i.e., a rather steep power-law, and for $\varepsilon_e = 0.125$ and $\varepsilon_B = 0.10$.

We reproduce the overall multi-frequency radio light behavior with a jet that has a half-opening angle $\theta_j \sim 6°$ (Table S7). The jet has an isotropic equivalent kinetic energy, $E_{iso}$, of $8 \times 10^{53}$ erg, corresponding to a beaming-corrected kinetic jet energy (*16*) $E_{jet} = E_{iso} (1 - \cos\theta_j)/2 \sim 2 \times 10^{51}$ erg. This energy is injected at a distance of $r_{inj} = 10^{16}$ cm from the central engine, and the jet is embedded in a circumnuclear ambient medium of a constant number density of $n_e \sim 3.7 \times 10^4$ cm$^{-3}$ up to a fiducial distance within the AGN torus of $r_f = 6.75 \times 10^{17}$ cm (Fig. S6A). This dense medium causes the jet to rapidly decelerate. While densities orders of magnitude higher can be found in the clouds of the broad line regions (BLR) surrounding the AGNs, the filling factors within these regions are very low (~$10^{-2}$).

Our modeling of the radio jet indicates that while it is initially relativistic ($\Gamma = 10$ and $\beta \approx 0.99$ at day zero, and $\Gamma = 1.8$ and $\beta \approx 0.87$ at ~1.1 days), as soon as ~16 days (in the observer frame) after the ejection the jet has already decelerated significantly ($\Gamma \approx 1.3$; $\beta \approx 0.64$). At a distance $r_f$, the jet reaches the edge of the torus and encounters an external medium with a 100 times lower density falling off then with distance as $(r/r_f)^{-2.5}$. The apparent kink seen in the radio light curves at all frequencies (Fig. 2A) happens also when the jet head crosses the outer boundary of the constant density medium, $r_f$. This distance corresponds to about 270 light days. Since the jet head crosses this boundary at about 1065 days after the ejection (as measured in the galaxy frame), the time at which this is observed in the light curves is roughly around day 795 (Figs. 2A and S6A). The sudden drop in density causes the jet to accelerate from $\beta \sim 0.09$ to $\beta \sim 0.22$ (Fig. S6A), converting a fraction of its internal energy into kinetic energy. The jet therefore reaches quasi-asymptotic values $\Gamma \approx 1.0$ and $\beta \approx 0.22$ by day ~800. We note that the actual model simulations were done for a 1-dimensional jet (with a viewing angle of 0°). However, as demonstrated in (*16*), the results for a larger viewing angle of $\theta_{obs}$ up to ~40°) are similar.

The model reproduces well both the early and late time radio emission from Arp 299-B AT1. The early rise at high frequencies implies the existence of significant external absorption close to the supermassive black hole of Arp 299-B1. This is due to free-free absorption of dense gas ($n_{e, torus} \sim 3.7 \times 10^4$ cm$^{-3}$) made of thermal electrons within the torus, with $T_e \sim 10^4$ K. The two earliest observations at 8.4 GHz, together with the late time radio observations, provide a stringent constraint on the distribution of $T_e$ within the torus. The early radio observations (t < 300 days) require that the electron temperature in the inner regions of the torus ($r \leq 6 \times 10^{17}$ cm) must be at least $T_e \sim 10^4$ K. On the other hand, $T_e$ must be around or lower than 3000 K close to the outer edge of the torus, providing the larger absorption required by the radio observations. If the temperature at the outer edge of the torus was of ~$10^4$ K instead, the parameters of the jet required to explain the first two epochs of 8.4 GHz observations would yield a much lower flux density for the late time (t ~ 400-500 days) radio emission than is actually observed.

## Constraints on the viewing angle and intrinsic velocity of the radio jet

Our VLBI observations provide constraints on the viewing angle, $\theta_{obs}$, and the average intrinsic speed, $\beta$, of the jet. We can constrain the allowed combinations using standard relations in the twin-relativistic jet model *(81)*, which assumes that the ejecta are double-sided and intrinsically the same (jet and the counter-jet have equal intrinsic emitting powers and equal velocities, $\beta$, in opposite directions), and that the interstellar medium which the jet and the counter-jet impact with is similar. The apparent jet speed, $\beta_{app}$, and the ratio between the flux densities of the jet and the counterjet, $R$, are then related with $\beta$ and $\theta_{obs}$ as follows:

$$\beta_{app} = \beta \sin\theta_{obs} / (1 - \beta \cos\theta_{obs}) \tag{S1}$$

and

$$R = [(1 + \beta \cos\theta_{obs})/(1 - \beta \cos\theta_{obs})]^{3-\alpha} \tag{S2}$$

where $\alpha$ is the spectral index, defined as $S_\nu \sim \nu^\alpha$, with $S_\nu$ the flux density at frequency $\nu$. Both the apparent speed of the jet, $\beta_{app}$, and the ratio of the flux density of the jet to the (undetected) counterjet, $R$, can thus be used to directly constrain $\theta_{obs}$ and $\beta$. We show in Fig. S7 the constraints on $\theta_{obs}$ and $\beta$, considering 3-$\sigma$ uncertainties for the measured proper motion, and a realistic 5-$\sigma$ upper limit for the counter-jet of Arp 299-B AT1 at late times (where $\sigma$ is the off-source r.m.s. noise within the B1 nucleus, and is typically about 30 µJy/b). The measured average value $\beta_{app}$ and its uncertainties delimit a region of allowed values of $\theta_{obs}$ and $\beta$ (solid lines in Fig. S7). In addition, the non-detection of the counterjet implies a curve corresponding to a constant value of $R$ (dashed curve in Fig. S7).

For the estimate of $R$, we have accounted for the free-free absorption from both the AGN torus and the external medium (as per Table S8), as well as for relativistic boosting, deboosting and light-travel-time effects. For values of the average intrinsic jet speed, $\beta$, larger than about 0.3, free-free absorption from the AGN torus and the external medium can be neglected, and the observed ratio $R$ coincides with the value predicted by Eq. (S2), which assumes no absorption. For smaller values of $\beta$, free-free absorption effects quickly become very significant, implying that a counter-jet at a relatively large angle (~25°-35°) would remain undetected. Note also that there is a minimum value for $\beta$. For values of $\beta$ lower than about 0.10, the deboosting is not high enough to compensate for the free-free absorption and produce a brightness ratio as high as the one observed.

In summary, our VLBI observations and radio modeling yield an average intrinsic jet speed of $\beta = 0.22\pm0.02$, and constrain the jet viewing angle to be within the range $\theta_{obs} \sim 25°$-35°. Viewing angles larger than ~50° are ruled out by the VLBI observations, which disfavors the radio jet being produced by normal, quiescent AGN accretion activity, where the jet would be expected to be launched nearly perpendicular to our line-of-sight to the AGN in Arp 299-B1, while it is seen nearly edge-on *(11)*.

## Origin of the IR emission

Inspection of the IR SED of Arp 299-B AT1 shows that it can be described well by a single blackbody component (see Fig. 3). We used the EMCEE python implementation *(82)* of the

Markov chain Monte Carlo (MCMC) method to estimate the best-fit blackbody parameters with 1σ confidence intervals for each epoch with 4.5 μm Spitzer photometry. For this we interpolated the ground-based *JHKs* and 3.6 μm Spitzer photometric measurements to the epochs of the 4.5 μm observations. The blackbody parameters are listed in Table S6, and their evolution is shown in Fig. S2, together with the blackbody luminosity and total radiated energy, obtained by integrating the luminosity over time assuming it changes linearly between the epochs. We note that this luminosity does not include the contribution of any cooler dust, which would radiate at longer wavelengths, hence this yields a lower limit to the total radiated energy of $1.5 \times 10^{52}$ erg.

We assume that the near- and mid-IR flux was due to thermal emission from dust heated by optical and UV emission from the transient event (e.g., *83*). Here, the transient is surrounded by local dust that absorbs and re-radiates most of the UV and optical photons emitted and radiates initially as a T ~ 1050 K blackbody. The blackbody radius indicates a size for the IR emitting region of ~0.04 pc initially, and later increasing to ~0.13 pc. The fact that the IR SED is well described by a blackbody up to 4.5 and 8.0 μm means that the dust shell local to the transient must be optically thick at least up to these wavelengths, corresponding to a visual extinction of $A_V > 20$ mag for a Galactic extinction law (*84-85*).

Given the high core-collapse SN rate of ~0.3 SN yr$^{-1}$ within the circumnuclear region of Arp 299-B1 (*8*), a SN is the most likely explanation for the source detected close to the nucleus B1 in the Mar. 2006 HST F814W filter observations (see Fig. S1). We emphasize that its contribution to the IR light curves is very small. Indeed, assuming a normal Type II SN (the most common core collapse SN type), we expect a total radiated energy less than a few times $10^{49}$ erg (*e.g.*, *86*) for this event. This is negligible compared to the total radiated energy of Arp 299-B AT1 event, if most of this energy was absorbed by dust and re-radiated in the IR. We do not see any evidence of this transient in our IR light curves of Arp 299-B AT1 (see Fig. S2). Furthermore, the optical HST images show a brightening of the nucleus from 2006 to 2010 corresponding to a new source of F814W = 19.8 (corresponding to 0.03 mJy) which is substantially brighter than extrapolated from our IR blackbody fits. For comparison the B1 nucleus brightened by ~0.8 mJy in *J*, ~5 mJy in *H* and ~16 mJy in *K*-band over the same period of time. Therefore the optical detection is not consistent with emission from hot dust. Instead, the *I*-band flux could be explained by optical emission of Arp 299-B AT1 scattered from free electrons above/below the broad line region (*87*) and/or from the polar dust.

For a transient event close to the SMBH in Arp 299-B1 we would expect a large extinction due to the AGN dusty torus (*10-11*), given the large column density of ~3 x $10^{24}$ cm$^2$ towards the accreting SMBH. The shape of the IR SED is very close to a single component blackbody in all the epochs, indicating a narrow range of dust temperatures. This is also not expected if there was substantial 'foreground' extinction due to the reddening altering the shape of the SED and the likely contribution of cooler dust. Therefore, absorption and re-radiation by dust in the polar regions of the AGN torus (*24,88*) appears to be the most plausible scenario for explaining the observed IR SED. In the surroundings of Arp 299-B1 we expect such polar dust clouds to suffer from a relatively low foreground extinction (*89*) in agreement with our single component blackbody fits to the IR SED of the transient. The presence of dust in the polar regions of at least some AGN, which is most likely concentrated in discrete optically thick clouds, is now a generally well accepted idea supported by mid-IR interferometry (*88*).

We have fitted the spectral energy distributions of the pre- and post-outburst components of Arp 299-B1 using the CYGNUS (*90*) radiative transfer code, which models the emission from a starburst and the AGN torus predicted by the standard unified model. The GYGNUS starburst model is described in (*91-92*) and has been used extensively for modeling the IR emission of both local and high-$z$ starburst galaxies. The model treats a starburst as an ensemble of giant molecular clouds (GMCs) centrally illuminated by recently formed stars, incorporates the stellar population synthesis model (*93*), and considers the effect of *transiently* heated grains/PAHs in the radiative transfer. An important feature of the model is that the GMCs are at different evolutionary stages. The key parameters of the model are the initial optical depth of the molecular clouds $\tau_V$, the age of the starburst $t_*$ and the e-folding time of the exponentially decaying star formation history.

The CYGNUS AGN model follows the prescription developed by (*94*) and can additionally include the effect of polar dust (*24, 95*). The model self-consistently solves the radiative transfer problem in the AGN torus to determine the temperature distribution of dust grains which have a range of sizes and composition. The model also self-consistently takes into account absorption of emission from the hot inner torus by dust in the cooler outer parts of the torus depending on the inclination. The model treats the AGN disc as a tapered one, i.e., the height of the disc increases with distance from the black hole, but tapers off to a constant height in its outer part. In this work we used a large grid of tapered discs (*94*). The parameters of the model are the ratio of outer to inner radius $r_2/r_1$ the equatorial UV optical depth $\tau_{UV}$, the opening angle of the disc $\theta_0$ and the inclination $i$.

The pre-outburst SED was adapted from (*9*) including also data points at near-IR wavelengths whereas the post-outburst SED was obtained by combining the former with the near- and mid-IR fluxes of the outburst at 734 days after its first mid-IR detection (corresponding to 54095.4 MJD) which is the last epoch with 8.0 μm photometry available. The spectral resolution of the Spitzer/IRS data was reduced so that they are better matched to the resolution of the radiative transfer models. For fitting the SEDs we used a grid of 6475 AGN torus templates, 360 starburst templates and a single polar dust template for each epoch. The SEDs are fitted with the MCMC SED fitting code SATMC (*96*) and after post-processing of the fitting data the covering factor of polar dust $f_c$ and its uncertainty are computed. SATMC combines and interpolates these grids to generate the required number of models to reach convergence. The models presented in Fig. 4 (for model parameters see Table S7) demonstrate that both the pre- and post-outburst SEDs can be modeled successfully adopting plausible parameters for the starburst, the AGN dusty torus, and the polar dust (incl. the opening angle, inclination, polar dust covering factor). We further assume that these parameters do not change as a result of the outburst, except for the temperature and luminosity of the polar dust that are left as a free parameter. In the post-outburst case $f_c$ = 23-78%. This implies that the observed radiated energy from the outburst must be divided by $f_c$ to give the total radiated energy i.e. ~1.9-6.5 x$10^{52}$ erg.

## Origin of the X-ray emission

The absence of significant variability in the soft X-ray band (0.3-8.0 keV), as given by pre- and post-burst *Chandra* and *XMM*-Newton observations, can be easily explained by the large column density (~3 x $10^{24}$ cm$^2$) towards the AGN in the nucleus B1: low-energy X-ray photons are in this case scattered multiple times, removing any evidence of an X-ray flare due to the different light travel delays introduced by the scattering.

Therefore, the only possibility of clearly detecting a sign of TDE emission at X-rays towards the nucleus of B1 would have been hard (> 10 keV) X-ray observations (*97*). While no hard X-ray observations were made shortly after the burst, the hard X-ray emission decreased by a factor of two between 2001 and 2013 *(10)*. Such X-ray variability is not uncommon in AGNs *(98)* but could be expected also in the case of a TDE, as the increase in the number of soft (optical) photons locally to the SMBH cools the corona, partially depleting the region from hard X-ray photons.

## The mass of the disrupted star

Based on our IR observations and modeling, the total radiated energy of Arp 299-B AT1 is between $1.9 \times 10^{52}$ erg and $6.5 \times 10^{52}$ erg. Following *(99)* we can estimate the maximum radiated energy by disruption of a star with a mass M.

$$E_{\rm rad,max} = \eta f_{\rm acc} M c^2 / 2 \sim 10^{52} (\eta/0.1) (f_{\rm acc}/0.1) (M/M_\odot) \text{ erg} \tag{S3}$$

where $f_{\rm acc}$ is the accreted fraction and $\eta$ is the radiative efficiency. Assuming $f_{\rm acc} = 0.1$ and $\eta = 0.1$, which are consistent with observations of optical TDEs *(99)*, we find that the disruption of a star in the range of 1.9 to 6.5 $M_\odot$ can explain the energetics of Arp 299-B AT1. Stars in this mass range can be disrupted by the $\sim 2 \times 10^7 M_\odot$ *(10)* SMBH in Arp 299-B1 *(24)*.

## Emitting and absorbing regions surrounding Arp 299-B AT1

In Fig. 4 (right panel), we present a schematic diagram showing the assumed geometry of the emitting and absorbing regions surrounding Arp 299-B AT1. The disruption of a star by the tidal forces of the SMBH in the nucleus Arp 299-B1 launches a relativistic jet that propagates outwards from the SMBH. The jet crosses the AGN torus, which is about 100 times denser than the surrounding circumnuclear medium. Most of the emission from the transient in the direction of the observer is thus absorbed and scattered by the dense gas and dust within the torus. However, a fraction of the photons can escape the torus in the polar direction and are observable in the near- and mid-IR range as thermal re-radiation by the polar dust.

Once the jet breaks out from the high-density region within the AGN torus, it expands freely and can be observed at radio wavelengths. The main source of opacity for the jet is external absorption. Consequently, the emission of the radio jet is detected first at high frequencies (Fig. S3; bottom right panel) and later at lower frequencies. The drop in radio flux density results from the period in which the jet drills its way through the very dense regions of the obscuring torus. The early rise of the 8.4 GHz flux density of Arp 299-B AT1, together with its apparent proper motion as measured with VLBI, imply that the distance travelled by the jet inside the high-density region of the torus is $l_t \sim 6.8 \times 10^{17}$ cm, or $\sim 0.22$ pc.

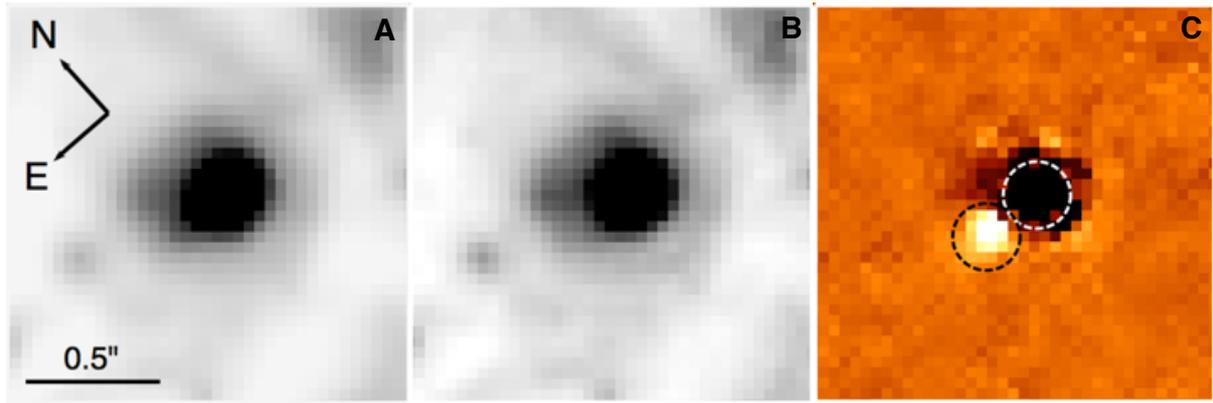

**Figure S1**. **HST images of the site of Arp 299-B AT1.** HST F814W (close to *I*-band) images of Arp 299-B AT1 taken on 2006 Mar. 19 (**A**; 2453813.8 JD) and 2010 Jun. 24 (**B**; 2455372.4 JD). (**C**) Difference image. The source 0.2″ to the East from the nucleus which appears white in the difference image was present in 2006 but not in 2010, while the center of the nucleus, which was brighter in 2010 appears as a dark source. The dashed circles indicate the 0.1″ apertures used for photometry.

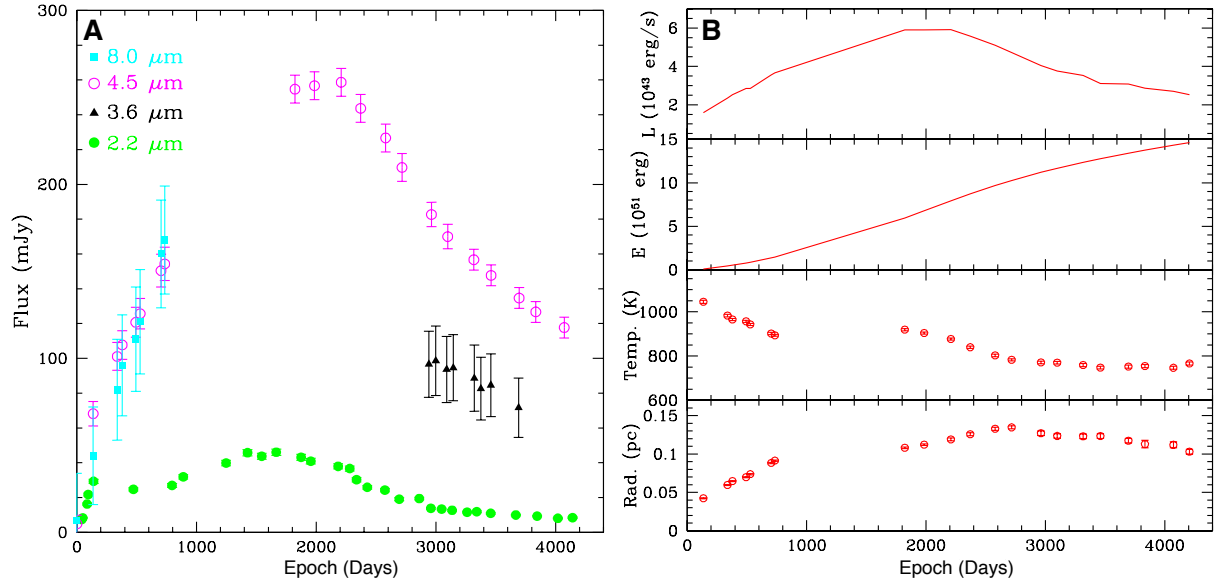

**Figure S2. IR observations and derived properties for Arp 299-B AT1.** (**A**) *K*-band (2.2 μm) and Spitzer 3.6, 4.5 and 8.0 μm light curves (after removing the quiescent nuclear flux). The epoch when the mid-IR light curves were observed to start rising (2004 Dec. 21.6 corresponding to 2453361.1 JD) has been adopted as day 0. (**B**) The evolution of the derived blackbody radius, temperature, luminosity and total energy over time. Our blackbody model fits (see Fig. 3) suggest that an expanding (from about 0.04 pc to 0.13 pc) and cooling (from about 1050 to 750K) source can account for the observed near- and mid-IR behavior. Integrating the blackbody luminosities to the current epoch yields a total radiated energy of about $1.5 \times 10^{52}$ erg.

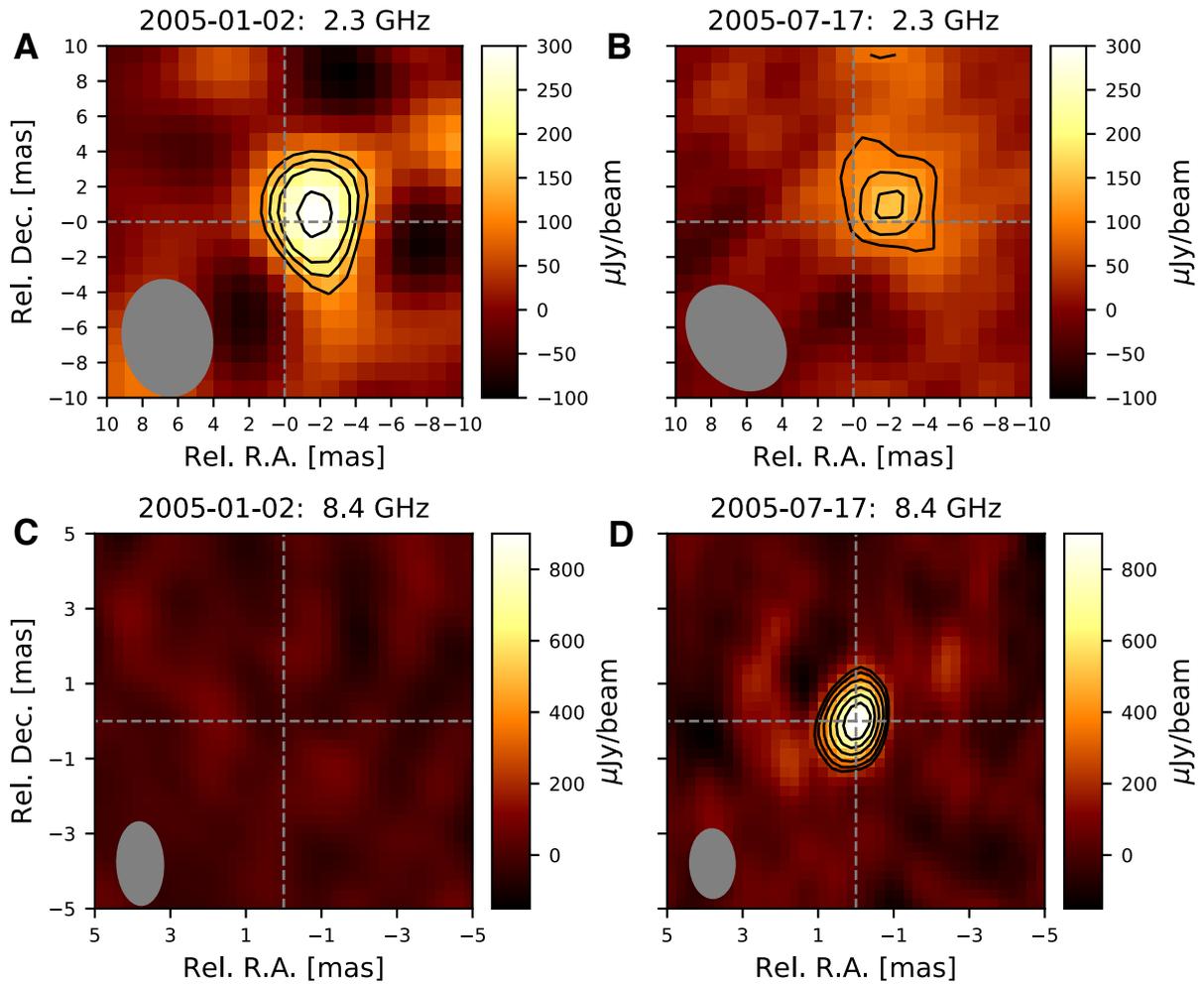

**Figure S3. The appearance of the transient Arp 299-B AT1 in the nucleus B1 at radio wavelengths.** Pre-discovery (2453372.5 JD) VLBI observations of Arp 299-B AT1 at 2.3 (**A**) and 8.4 GHz (**C**). Pre-existing radio emission is seen at 2.3 GHz, which we identify with the core of the AGN in Arp 299-B1. Post-discovery (2453568.5 JD) VLBI observations at 2.3 (**B**) and 8.4 GHz (**D**), showing the appearance of Arp 299-B AT1 at high radio frequencies (dashed vertical and horizontal lines are centered at the peak position of 8.4 GHz emission). In each map, the contours are drawn at (3,4,5,7,10,15) times the off-source r.m.s. All maps are centered at RA = 11h28m30.9875529s, Dec = 58°33′ 40.783601″ (J2000.0).

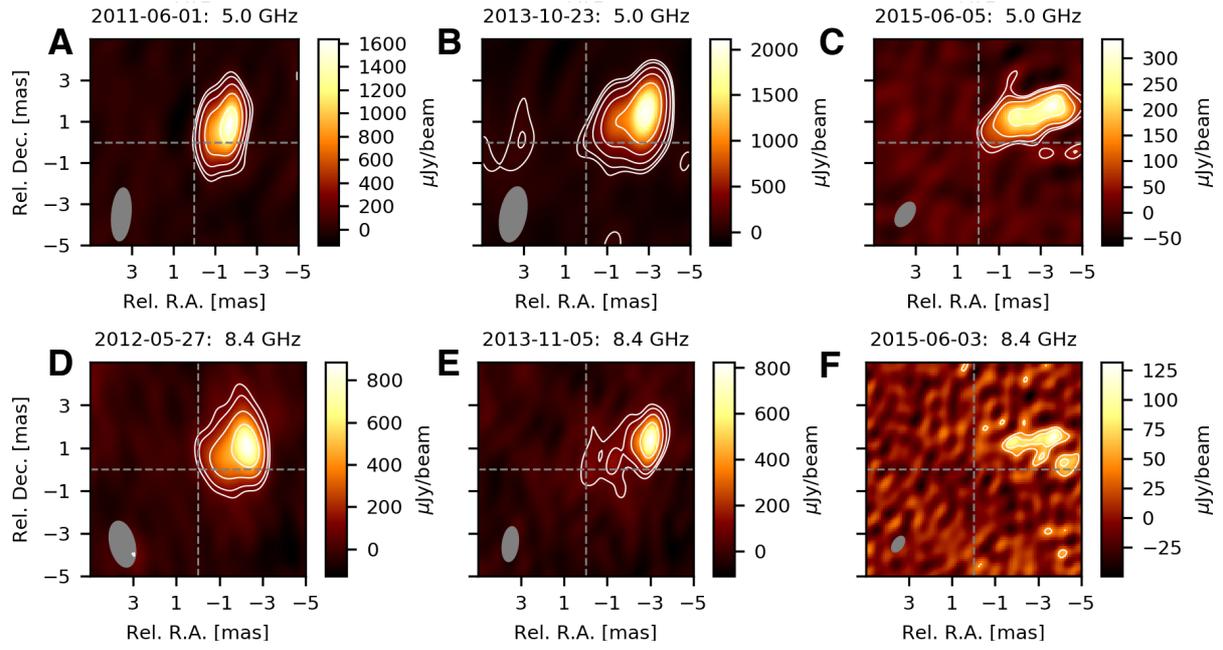

**Figure S4. The radio jet associated with Arp 299-B AT1.** VLBI observations of Arp 299-B AT1 at 5.0 GHz on 2011 Jun. 1 (**A**; 2455713.5, JD), 2013 Oct. 23 (**B**; 2456588.5 JD), and 2015 Jun. 5 (**C**; 2457178.5 JD) and at 8.4 GHz on 2012 May 27 (**D**; 2456074.5 JD), 2013 Nov. 5 (**E**; 2456601.5 JD), and 2015 Jun. 3 (**F**; 2457176.5 JD). There is evidence for resolved emission at 5.0 GHz in Jun. 2011 (**A**), which is confirmed by the subsequent 5.0 and 8.4 GHz observations (**B** through **F**). Data are displayed in the same way as in Fig. S3.

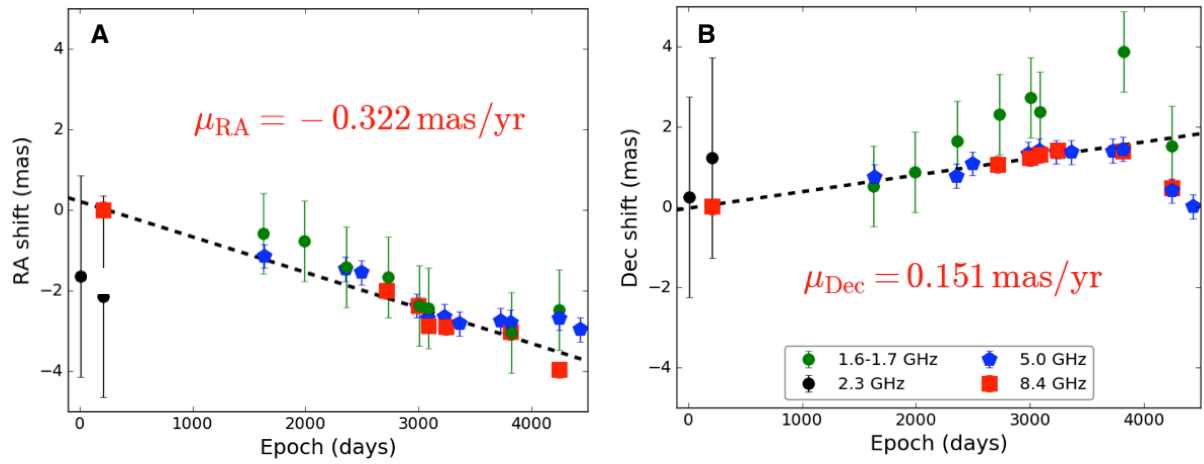

**Fig. S5. Proper motion of Arp 299-B AT1 as measured from VLBI observations.** Angular shift in right ascension (RA) **(A)** and declination (Dec) **(B)**, as function of time. The epoch when the mid-IR light curves were observed to start rising (2004 Dec. 21.6 corresponding to 2453361.1 JD) has been adopted as day 0. The data show that the head of the jet consistently moved to the North-West for the first ~4000 days at an average apparent speed of $(0.25 \pm 0.03)\,c$.

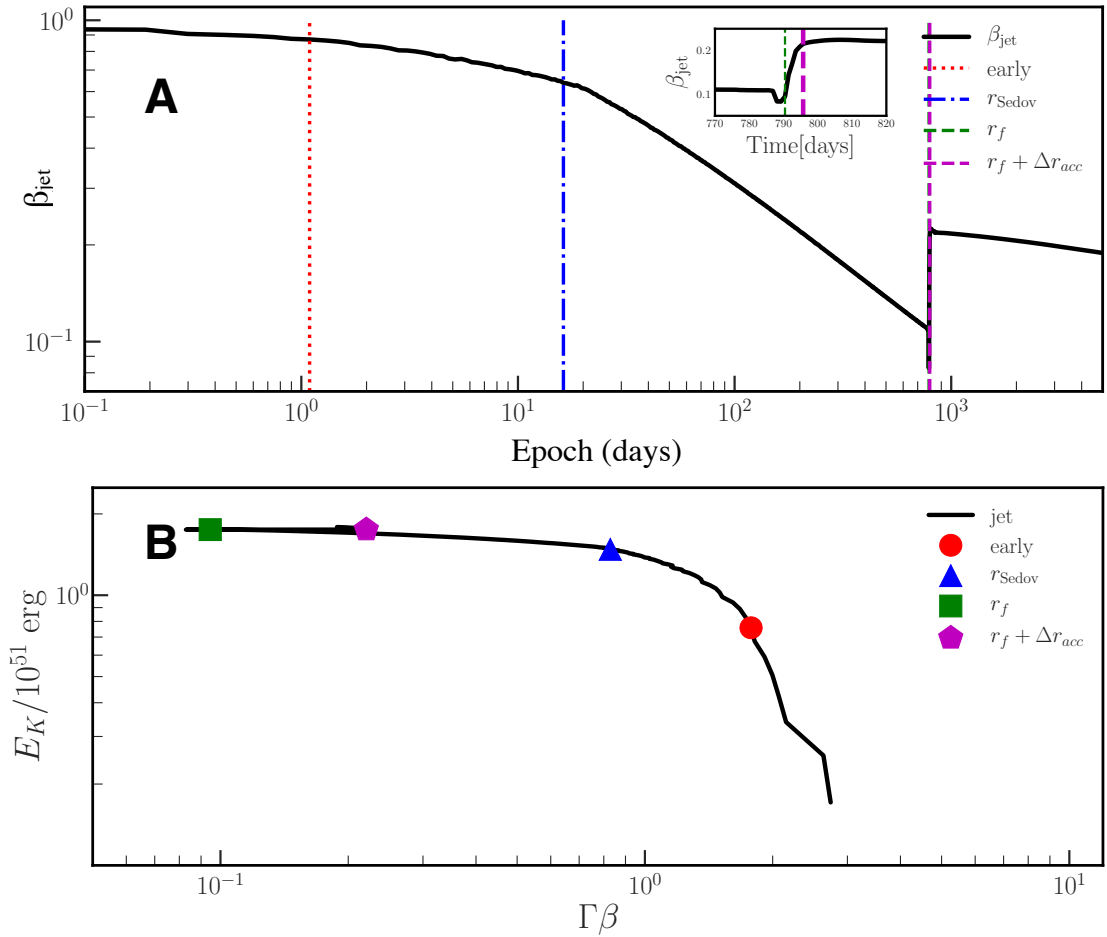

**Figure S6. (A) Model speed of the TDE jet in Arp 299-B AT1 vs. time (solid line).** The dotted, dashed and dot-dashed lines highlight relevant times for the evolution of the jet speed: early time (at about 1 day in the observer frame, dotted line), Sedov time (the time at which the rest mass energy of the accumulated material equals the jet $E_{iso}$, dot-dashed line), and the epochs at which the jet reaches the lowest velocity (close to the outer edge of the AGN torus; thin dashed line in the inset figure) and the acceleration (very shortly after leaving the dense torus and entering the much less dense circumnuclear material; thick dashed line in the inset figure). **(B)** Modeled (beaming-corrected) jet kinetic energy versus speed of the jet (as in Fig. 2), highlighting the same moments as in panel A.

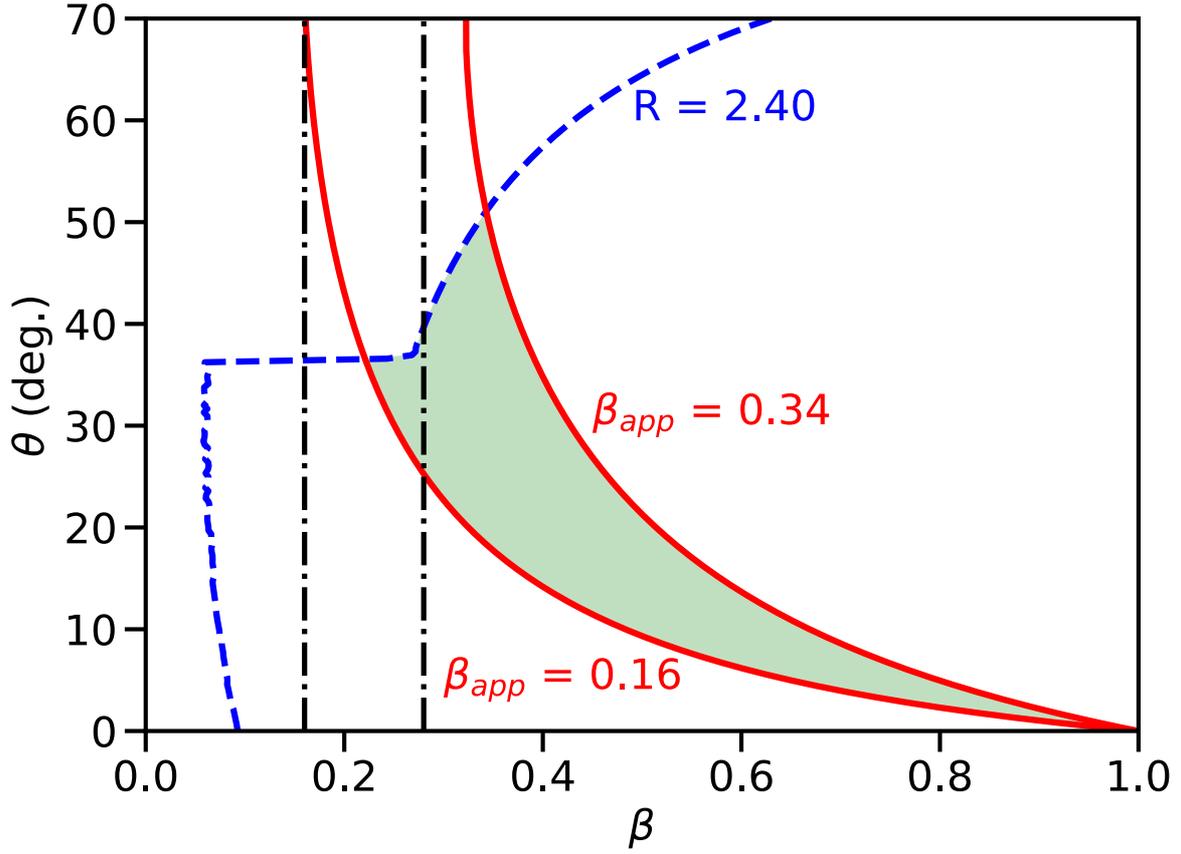

**Figure S7. Viewing angle versus average intrinsic jet speed ($\theta_{obs}$, $\beta$) parameter space for Arp 299-B AT1.** The shaded area below the dashed line corresponds to the allowed range of viewing angles and average intrinsic speeds for the jet, as constrained by our VLBI observations. The solid lines correspond to the 3-σ uncertainties for the measured proper motion of the jet. The dashed line corresponds to the 5-σ upper limit for the counter-jet of Arp 299-B AT1 at late times. The dash-dotted lines correspond to the 3-σ range of values of the intrinsic jet speed found from our radio modeling ($\beta$ = 0.22 ± 0.02). Our VLBI observations and radio modeling yield an average intrinsic jet speed of $\beta \sim$ (0.2-0.3) corresponding to a viewing angle of $\theta_{obs} \sim$ (25°-35°).

**Table S1. Log of ground-based near-IR photometry.** Col. 1: Julian Date since day 2400000; col. 2: the telescope and instrument used; col. 3: the flux densities given after subtraction of the quiescent nuclear flux. The photometric errors are dominated by the systematic uncertainties and are assumed to be 4%.

| Julian Date (2400000+) | Telescope/instrument | Flux density $J$-band (mJy) | Flux density $H$-band (mJy) | Flux density $K$-band (mJy) |
|---|---|---|---|---|
| 50875.0 | IRTF | - | - | - |
| 53162.5 | WHT/LIRIS | - | - | - |
| 53400.8 | WHT/LIRIS | - | - | 7.36 |
| 53410.1 | UKIRT/UFTI | 1.22 | 3.43 | 8.33 |
| 53446.7 | WHT/LIRIS | - | - | 16.27 |
| 53455.6 | NOT/NOTCam | - | - | 21.72 |
| 53500.7 | TIFCAM | 3.58 | - | 29.33 |
| 53833.7 | NOT/NOTCam | - | 7.56 | 24.75 |
| 54156.6 | NOT/NOTCam | 2.59 | 8.61 | 26.95 |
| 54250.4 | NOT/NOTCam | - | - | 31.92 |
| 54608.4 | NOT/NOTCam | - | 12.15 | 39.82 |
| 54786.7 | NOT/NOTCam | - | 14.29 | 45.76 |
| 54905.5 | NOT/NOTCam | - | - | 43.76 |
| 55027.5 | NOT/NOTCam | 4.20 | 13.70 | 46.04 |
| 55234.8 | NOT/NOTCam | - | - | 43.11 |
| 55315.5 | NOT/NOTCam | 4.03 | 12.97 | 40.89 |
| 55543.7 | NOT/NOTCam | 3.46 | 11.07 | 37.95 |
| 55641.4 | NOT/NOTCam | - | - | 36.66 |
| 55697.5 | NOT/NOTCam | 2.52 | 8.35 | 30.25 |
| 55787.4 | NOT/NOTCam | - | - | 25.88 |
| 55934.7 | NOT/NOTCam | - | 6.46 | 24.30 |
| 56054.5 | NOT/NOTCam | 2.16 | 4.27 | 19.05 |
| 56221.8 | NOT/NOTCam | - | - | 19.41 |
| 56318.7 | NOT/NOTCam | - | 4.08 | 13.79 |
| 56408.5 | NOT/NOTCam | - | 2.96 | 13.39 |
| 56496.4 | NOT/NOTCam | - | 3.45 | 12.68 |
| 56621.7 | NOT/NOTCam | - | - | 11.59 |
| 56701.6 | NOT/NOTCam | - | - | 11.85 |
| 56818.6 | NOT/NOTCam | 1.88 | 3.01 | 10.91 |
| 57028.6 | NOT/NOTCam | - | - | 9.97 |
| 57206.4 | NOT/NOTCam | - | - | 9.33 |
| 57379.6 | NOT/NOTCam | - | - | 8.07 |
| 57502.5 | NOT/NOTCam | 1.42 | 2.28 | 8.45 |
| 57561.4 | NOT/NOTCam | - | - | 9.08 |

**Table S2. Log of high-resolution near-IR imaging observations.** Col. 1: Julian Date since day 2400000; col. 2-3: the telescope, instrument and filter used; col. 4-5: the root-mean-square (rms) uncertainty of the derived geometric transformation between each Gemini image and the reference HST image in right ascension (RA) and declination (Dec); cols. 6-7: the offset between the position of Arp 299-B AT1 (as measured in the difference images) and the B1 nucleus (as measured in the reference HST image) in RA and Dec. In RA Arp 299-B AT1 is found to be coincident with the quiescent nucleus in all epochs within 1σ. In Dec we find a maximum offset of 37 milliarcseconds (at about 2σ level).

| Julian Date (2400000+) | Telescope/ Instrument | Filter | rms in RA (mas) | rms in Dec (mas) | Offset in RA (mas) | Offset in Dec (mas) |
|---|---|---|---|---|---|---|
| 50757.5 | HST/NICMOS2 | $F222M$ | - | - | - | - |
| 54165.9 | Gemini/NIRI | $K$ | 17 | 13 | -7 | -22 |
| 54639.8 | Gemini/NIRI | $K$ | 17 | 14 | -3 | -23 |
| 54818.1 | Gemini/NIRI | $K$ | 18 | 15 | -10 | -23 |
| 54928.8 | Gemini/NIRI | $K$ | 19 | 16 | -6 | -23 |
| 54958.8 | Gemini/NIRI | $K$ | 19 | 15 | -6 | -28 |
| 54859.1 | Gemini/NIRI | $K$ | 17 | 16 | -12 | -37 |
| 55252.0 | Gemini/NIRI | $K$ | 18 | 16 | -20 | -37 |
| 55317.8 | Gemini/NIRI | $K$ | 19 | 15 | -5 | -27 |
| 55668.8 | Gemini/NIRI | $K$ | 18 | 16 | -14 | -2 |
| 55956.1 | Gemini/NIRI | $K$ | 14 | 12 | -4 | -11 |

**Table S3. Log of Spitzer observations.** Col. 1: Julian Date since day 2400000; col. 2-4: the flux densities are listed before removing the quiescent nuclear flux (96.4, 164.3 and 879 mJy at 3.6, 4.5 and 8.0 μm, respectively) and the 1σ uncertainties are listed in parentheses. The epoch when the mid-IR light curves were observed to start rising (2004 Dec. 21.6 corresponding to 2453361.1 JD) has been adopted as day 0 in Figures 2, 3 and S2.

| Julian Date (2400000+) | Flux density 3.6μm (mJy) | Flux density 4.5μm (mJy) | Flux density 8.0μm (mJy) |
|---|---|---|---|
| 52991.7 | 96.4(2.9) | 165.1(5.0) | 875(26) |
| 53152.2 | - | 162.7(4.9) | 882(26) |
| 53328.3 | - | 165.2(5.0) | 881(26) |
| 53361.1 | - | 169.1(5.1) | 886(27) |
| 53497.2 | - | 232.5(7.0) | 923(28) |
| 53698.6 | - | 265.4(8.0) | 961(29) |
| 53740.3 | - | 272.0(8.2) | 975(29) |
| 53854.0 | - | 285.0(8.6) | 990(30) |
| 53889.9 | - | 290.0(8.7) | 1000(30) |
| 54064.6 | - | 314.7(9.4) | 1039(31) |
| 54095.4 | - | 318.6(9.6) | 1047(31) |
| 55183.5 | - | 419(8) | - |
| 55346.4 | - | 421(8) | - |
| 55569.5 | - | 423(8) | - |
| 55732.2 | - | 408(8) | - |
| 55940.9 | - | 391(8) | - |
| 56078.1 | - | 374(8) | - |
| 56302.7 | 193(19) | - | - |
| 56324.0 | - | 347(7) | - |
| 56360.3 | 195(20) | - | - |
| 56450.7 | 190(19) | - | - |
| 56460.1 | - | 334.3(7) | - |
| 56505.7 | 191(19) | - | - |
| 56677.9 | - | 321(6) | - |
| 56682.7 | 185(19) | - | - |
| 56737.4 | 179(18) | - | - |
| 56820.7 | 181(18) | - | - |
| 56822.1 | - | 312(6) | - |
| 57052.0 | 168(17) | - | - |
| 57056.7 | - | 299(6) | - |
| 57195.4 | - | 291(6) | - |
| 57433.4 | - | 282(6) | - |
| 57568.3 | - | 276(6) | - |

**Table S4. Log of interferometric radio observations for the source Arp 299-B AT1.** Col. 1: Julian Date since day 2400000; col. 2: the observing array (EVN=European VLBI Network, VLBA=Very Long Baseline Interferometry, VLA = Very Large Array, GBT = Green Bank Telescope); col. 3: observing frequency, in GHz; col. 4: Size of the radio interferometric synthesized beam at each epoch and for the given array, in (mas × mas); col. 5: peak flux density, in µJy beam$^{-1}$; col. 6: total flux density in µJy; col. 7: flux density uncertainty, in µJy. The values in cols. 5 and 6 for the VLA array have been obtained after having subtracted a baseline flux density at each frequency, following *(17)*. The quoted flux density uncertainties in col. 7 are 1σ when the source is detected, and include the sum in quadrature of the off-source r.m.s. noise in the image plus the calibration uncertainties. When the source was not detected, col. 7 gives the 3σ upper limit, where 1σ corresponds to the off-source r.m.s. noise of the image. 2004 Dec. 21.6 was adopted as day 0, corresponding to the Julian Date (JD) 2453361.1.

| Julian Date (2400000+) | Array | Freq. (GHz) | Beam size (mas × mas) | Peak flux density (µJy beam$^{-1}$) | Total flux density (µJy) | Flux density uncertainty (µJy) |
|---|---|---|---|---|---|---|
| 53311.5 | VLA | 8.4 | 780×730 | 0 | 0 | 390 |
| 53372.5 | VLBA+GBT | 8.4 | 2.2×1.2 | 0 | 0 | 99 |
| 53372.5 | VLBA+GBT | 2.3 | 6.6×4.0 | 0 | 0 | 195 |
| 53409.5 | VLA | 22.5 | 296×102 | 2704 | 2704 | 820 |
| 53539.5 | VLA | 22.5 | 850×300 | 11395 | 11395 | 3450 |
| 53568.5 | VLBA+GBT | 8.4 | 1.8×1.2 | 966 | 1150 | 132 |
| 53568.5 | VLBA+GBT | 2.3 | 6.5×4.8 | 0 | 0 | 210 |
| 53840.5 | VLA | 8.4 | 780×730 | 7670 | 7670 | 2320 |
| 54989.5 | EVN | 1.7 | 11.1×8.2 | 10161 | 9574 | 963 |
| 54994.5 | EVN | 5.0 | 2.7×1.1 | 5299 | 7645 | 767 |
| 55344.5 | EVN | 5.0 | 3.0×1.0 | 2124 | 3967 | 416 |
| 55352.5 | EVN | 1.7 | 5.9×2.3 | 13288 | 15143 | 1522 |
| 55713.5 | EVN | 5.0 | 2.6×0.9 | 1791 | 3428 | 361 |
| 55724.5 | EVN | 1.7 | 8.7×4.0 | 8195 | 8893 | 892 |
| 55855.5 | EVN | 5.0 | 3.2×1.7 | 2931 | 3815 | 391 |
| 56074.5 | EVN | 8.4 | 2.1×1.1 | 697 | 1557 | 171 |
| 56090.5 | EVN | 1.7 | 10.0×3.8 | 9037 | 9547 | 959 |
| 56231.5 | EVN | 5.0 | 2.4×0.8 | 4653 | 4653 | 489 |
| 56348.5 | EVN | 5.0 | 3.2×1.4 | 2149 | 3187 | 337 |
| 56358.5 | EVN | 8.4 | 1.7×0.8 | 767 | 2416 | 263 |
| 56364.5 | EVN | 1.7 | 6.4×3.0 | 6483 | 9826 | 985 |
| 56438.5 | EVN | 5.0 | 2.8×0.9 | 2150 | 3078 | 312 |
| 56449.5 | EVN | 1.7 | 6.9×3.3 | 6161 | 9804 | 983 |
| 56452.5 | EVN | 8.4 | 1.4×0.5 | 680 | 770 | 107 |
| 56588.5 | EVN | 5.0 | 2.6×1.2 | 1306 | 3010 | 307 |
| 56601.5 | EVN | 8.4 | 1.6×0.7 | 760 | 1349 | 146 |
| 56720.5 | EVN | 5.0 | 2.7×1.0 | 1899 | 2204 | 224 |
| 57085.5 | EVN | 5.0 | 3.5×1.4 | 738 | 1520 | 179 |
| 57176.5 | EVN+VLBA | 8.4 | 0.7×0.4 | 95 | 878 | 153 |
| 57178.5 | EVN+VLBA | 5.0 | 1.3×0.7 | 486 | 2721 | 316 |
| 57185.5 | EVN+VLBA | 1.7 | 5.1×2.7 | 2443 | 5913 | 603 |

| 57605.5 | VLBA | 5.0 | 5.2×1.5 | 1152 | 1841 | 201 |
| 57605.5 | VLBA | 1.5 | 13.7×4.3 | 4501 | 5813 | 597 |
| 57605.5 | VLBA | 8.4 | 2.3×0.7 | 274 | 562 | 114 |
| 57788.5 | VLBA | 5.0 | 5.1×1.5 | 774 | 1880 | 200 |

**Table S5. Log of soft X-ray observations.** Col. 1: Julian Date since day 2400000; col. 2: name of the observatory and identificator of the specific observation; col. 3: full observed flux in the 0.3-8 keV energy band, in $10^{-13}$ erg s$^{-1}$ cm$^{-2}$; col. 4: model flux of the power-law spectral component n the 0.3-8 keV energy band corrected for the photoelectric absorption, in $10^{-12}$ erg s$^{-1}$ cm$^{-2}$; col. 4: photon index of the power-law spectral component; col. 4: equivalent width of the iron 6.4-keV fluorescent emission line, keV.

| Julian Date (2400000+) | Instrument, ObsId | Full flux 0.3-8 keV ($10^{-13}$ erg cm$^2$ s$^{-1}$) | Power-law flux 0.3-8 keV, unabsorbed ($10^{-12}$ erg cm$^2$ s$^{-1}$) | Power-law photon index | Fe K Equivalent Width (keV) |
|---|---|---|---|---|---|
| 52035.5 | *XMM-Newton*, ID 0112810101 | $7.16^{+0.22}_{-0.13}$ | $14.90^{+0.70}_{-0.64}$ | $1.87^{+0.08}_{-0.08}$ | $-0.64^{+0.27}_{-0.28}$ |
| 55910.5 | *XMM-Newton*, ID 0679381101 | $8.26^{+0.38}_{-0.25}$ | $16.35^{+0.90}_{-0.90}$ | $1.82^{+0.09}_{-0.09}$ | $-0.31^{+0.18}_{-0.28}$ |
| 52103.5 | *Chandra*, ObsId 1641 | $1.84^{+0.19}_{-0.18}$ | $2.53^{+0.31}_{-0.28}$ | $1.10^{+0.26}_{-0.27}$ | $-0.31^{+0.23}_{-0.30}$ |
| 53415.5 | *Chandra*, ObsId 6227 | $1.48^{+0.27}_{-0.26}$ | $1.76^{+0.76}_{-0.35}$ | $0.99^{+0.56}_{-0.64}$ | $-0.59^{+0.48}_{-0.67}$ |
| 56363-56364 | *Chandra*, ObsId 15077+15619 | $2.13^{+0.10}_{-0.10}$ | $3.17^{+0.19}_{-0.17}$ | $1.36^{+0.11}_{-0.11}$ | $-0.75^{+0.16}_{-0.15}$ |

**Table S6. Derived Infrared properties of Arp 299-B AT1**. Col. 1: Julian Date since day 2400000; col. 2-3: blackbody parameters derived for the dates with 4.5μm photometry (the 1σ uncertainties in the last digit are listed in parentheses); col. 4: the blackbody luminosity; col. 5: the total radiated energy obtained by integrating the luminosity over time assuming it changes linearly between the epochs.

| Julian Date (2400000+) | Radius (pc) | Temperature (K) | Luminosity ($10^{43}$ erg s$^{-1}$) | Energy ($10^{51}$ erg) |
|---|---|---|---|---|
| 53497.2 | 0.0423(6) | 1045(7) | 1.54 | 0.091 |
| 53698.6 | 0.0596(7) | 983(6) | 2.35 | 0.434 |
| 53740.3 | 0.0646(7) | 965(5) | 2.53 | 0.522 |
| 53854.0 | 0.0699(7) | 957(5) | 2.85 | 0.786 |
| 53889.9 | 0.0738(8) | 943(5) | 2.85 | 0.874 |
| 54064.6 | 0.0883(9) | 902(5) | 3.56 | 1.36 |
| 54095.4 | 0.0914(9) | 894(4) | 3.66 | 1.45 |
| 55183.5 | 0.108(1) | 919(5) | 5.90 | 5.95 |
| 55346.4 | 0.112(1) | 904(5) | 5.90 | 6.78 |
| 55569.5 | 0.119(2) | 877(5) | 5.92 | 7.92 |
| 55732.2 | 0.126(2) | 840(5) | 5.57 | 8.73 |
| 55940.9 | 0.133(2) | 803(6) | 5.10 | 9.69 |
| 56078.1 | 0.135(2) | 783(5) | 4.72 | 10.3 |
| 56324.0 | 0.127(3) | 771(6) | 4.05 | 11.2 |
| 56460.1 | 0.124(3) | 770(6) | 3.76 | 11.7 |
| 56677.9 | 0.123(3) | 759(7) | 3.53 | 12.3 |
| 56822.1 | 0.123(3) | 748(6) | 3.11 | 12.8 |
| 57056.7 | 0.117(3) | 752(7) | 3.08 | 13.4 |
| 57195.4 | 0.113(5) | 755(7) | 2.87 | 13.7 |
| 57433.4 | 0.112(4) | 747(7) | 2.70 | 14.3 |
| 57568.3 | 0.103(3) | 766(7) | 2.53 | 14.6 |

**Table S7. Pre- and post-outburst properties of the nucleus B in Arp 299.** The luminosities of the AGN torus ($L_{AGN}$), starburst ($L_{SB}$), and polar dust ($L_{polar}$) components, the AGN fraction ($f_{AGN}$) and the polar dust temperature ($T_{polar}$) and covering factor ($f_C$) obtained from the infrared SED fits for Arp 299-B before and after (734 days after the first mid-IR detection corresponding to 2454095.4 JD) the outburst.

| | Pre-outburst | Post-outburst |
|---|---|---|
| $L_{AGN}$ ($10^{10}$ L☉) | $3.18^{+0.77}_{-0.87}$ | $3.02^{+1.31}_{-0.66}$ |
| $L_{SB}$ ($10^{10}$ L☉) | $9.84^{+1.15}_{-0.79}$ | $9.91^{+0.07}_{-0.16}$ |
| $L_{polar}$ ($10^{9}$ L☉) | $0.52^{+4.68}_{-0.52}$ | $7.78^{+4.05}_{-4.07}$ |
| $f_{AGN}$ | $0.24^{+0.04}_{-0.06}$ | $0.23^{+0.07}_{-0.04}$ |
| $T_{polar}$ (K) | 500 | 900 |
| $f_C$ | $0.03^{+0.40}_{-0.03}$ | $0.52^{+0.26}_{-0.29}$ |

**Table S8. Best-fit parameters of the TDE jet model for the radio light curves of Arp 299-B AT1.** The jet (col. 1) has a beaming-corrected kinetic energy, $E_{jet}$, and a (constant) half-opening angle $\theta_j$, and its bulk motion is described by the Lorentz factor, $\Gamma$; $p$ is the index of the injected power-law of relativistic electrons, $\varepsilon_e$ and $\varepsilon_B$ are the fractions of post-shock thermal energy going to relativistic electrons and magnetic fields, respectively, and $\zeta_e$ is the fraction of electrons being accelerated at the shock; $r_{inj}$ is the distance from the central engine at which $E_{jet}$ is injected; $n_e$ and $T_e$ are the thermal electron density and electron temperature inside the AGN torus (col. 2) and in the external medium (col. 3).

| Jet | Torus | External medium |
|---|---|---|
| $E_{jet} = 1.8 \times 10^{51}$ erg | $n_e (r < r_f) = 3.7 \times 10^4$ cm$^{-3}$ | $n_e (r > r_f) = 370 \, (r/r_f)^{-2.5}$ cm$^{-3}$ |
| $\theta_j = 6$ degrees | $T_e (r < r_i) = 10^4$ K | $T_e (r > r_f) = 3 \times 10^3$ K |
| $\Gamma$ (initial) = 10 | $T_e (r_i < r < r_f) = 3 \times 10^3$ K | $r_f = 6.75 \times 10^{17}$ cm |
| $p = 3.0$ | $r_i = 6.0 \times 10^{17}$ cm | |
| $\varepsilon_e = 0.125$ | $r_f = 6.75 \times 10^{17}$ cm | |
| $\varepsilon_B = 0.1$ | | |
| $\zeta_e = 1$ | | |
| $r_{inj} = 10^{16}$ cm | | |